\documentclass[aps,prb,superscriptaddress,twocolumn]{revtex4-1}
\pdfoutput=1
\usepackage{amsmath}
\usepackage{ amssymb }
\usepackage{graphicx}
\usepackage{color}

\begin{document}

\newcommand{\beq}{\begin{equation}}
\newcommand{\eeq}{\end{equation}}
\newcommand{\beqa}{\begin{eqnarray}}
\newcommand{\eeqa}{\end{eqnarray}}
\newcommand{\ben}{\begin{enumerate}}
\newcommand{\een}{\end{enumerate}}
\newcommand{\hs}{\hspace{0.5cm}}
\newcommand{\vs}{\vspace{0.5cm}}
\newcommand{\note}[1]{{\color{red} [#1]}}
\newcommand{\tim}{$\times$}
\newcommand{\bigo}{\mathcal{O}}
\newcommand{\bra}[1]{\ensuremath{\langle#1|}}
\newcommand{\ket}[1]{\ensuremath{|#1\rangle}}
\newcommand{\bracket}[2]{\ensuremath{\langle#1|#2\rangle}}
\renewcommand{\vec}[1]{\textbf{#1}}

\newcommand{\dagga}{{\phantom{\dagger}}}
\newcommand{\ud}{\,\mathrm{d}}

\newcommand{\bo}{B} 
\newcommand{\wkz}{\omega_{\vec{k},z}}
\newcommand{\qq}{q} 
\newcommand{\jp}{J'} 
\newcommand{\jpl}{J} 
\newcommand{\gk}{\gamma_{\vec{k}}}

\newcommand{\todo}[1]{{\color{red} {\bf [ToDo: #1]}}}


\title{Dynamical structure factors and excitation modes of the bilayer Heisenberg model}

\author{M. Loh\"ofer}
\affiliation{Institut f\"ur Theoretische Festk\"orperphysik, JARA-FIT and JARA-HPC, RWTH Aachen University, 52056 Aachen, Germany}

\author{T. Coletta}
\affiliation{School of Engineering, University of Applied Sciences of Western Switzerland, 1951 Sion, Switzerland}

\author{D. G. Joshi}
\affiliation{Institut f\"ur Theoretische Physik, Technische Universit\"at Dresden, 01062 Dresden, Germany}

\author{F. F. Assaad}
\affiliation{Institut f\"ur Theoretische Physik und Astrophysik, Universit\"at W\"urzburg, Am Hubland, 97074 W\"urzburg, Germany}

\author{M. Vojta}
\affiliation{Institut f\"ur Theoretische Physik, Technische Universit\"at Dresden, 01062 Dresden, Germany}

\author{S. Wessel}
\affiliation{Institut f\"ur Theoretische Festk\"orperphysik, JARA-FIT and JARA-HPC, RWTH Aachen University, 52056 Aachen, Germany}

\author{F. Mila}
\affiliation{Institute of Theoretical Physics, Ecole Polytechnique F\'ed\'erale de Lausanne (EPFL), 1015 Lausanne, Switzerland}
\date{\today}

\begin{abstract}

Using quantum Monte Carlo simulations along with higher-order spin-wave theory, bond-operator and strong-coupling expansions, we analyse the dynamical spin structure factor of the spin-half Heisenberg model on the square-lattice bilayer. We identify distinct contributions from the low-energy Goldstone modes in the magnetically ordered phase and the gapped triplon modes in the quantum disordered phase. In the antisymmetric (with respect to layer inversion) channel, the dynamical spin structure factor exhibits a continuous evolution of spectral features across the quantum phase transition, connecting the two types of modes. Instead, in the symmetric channel we find a depletion of the spectral weight when moving from the ordered to the disordered phase.
While the dynamical spin structure factor does not exhibit a well-defined distinct contribution from the amplitude (or Higgs) mode in the ordered phase, we identify an only marginally-damped amplitude mode in the dynamical singlet structure factor, obtained from interlayer bond correlations, in the vicinity of the quantum critical point.
These findings provide quantitative information in direct relation to possible neutron or light scattering experiments in a fundamental two-dimensional quantum-critical spin system.

\end{abstract}

\maketitle


\section{Introduction}

Advances in both energy and angular resolution of scattering experiments enable refined experimental studies of  collective excitations in strongly correlated quantum many-body systems. A prominent example is the three-dimensional quantum spin-dimer system TlCuCl${}_3$ where spin excitations have been mapped out in great detail using inelastic neutron scattering~\cite{Ruegg08,Ruegg14}. This compound exhibits a pressure-tuned zero-temperature transition from a gapped quantum disordered phase to an antiferromagnetically ordered phase. In addition to identifying the low-energy (transverse) Goldstone modes that accompany the spontaneous breaking of spin-rotation symmetry in the ordered phase, neutron scattering also detected the gapped (longitudinal) amplitude mode of the order-parameter field, frequently referred to as Higgs mode.
In the three-dimensional compound TlCuCl${}_3$, a rather successful quantitative theoretical account of these various spin excitation modes and the experimentally determined spectral weight can be obtained within a bond-operator mean-field description. In fact, the critical theory describing the underlying quantum critical point is the classical O(3) field theory in four dimensions, which exhibits only logarithmic corrections to a Gaussian fixed point. This fact also implies that for nearly critical three-dimensional collinear antiferromagnets, the amplitude mode becomes increasingly sharp upon approaching the quantum critical point~\cite{Affleck92}.

Upon lowering the dimensionality of the quantum magnet, however, the effects of both thermal and quantum fluctuations get significantly enhanced. In particular, for two-dimensional Heisenberg systems, true long-range order is restricted to zero temperature, and the quantum phase transition of spin-dimer models is controlled by the classical Wilson-Fisher fixed point in three dimensions where interactions are relevant \cite{fritz_note}. These interactions influence the visibility of the amplitude mode, as it can efficiently decay into pairs of Goldstone bosons. This has been examined intensively in recent years, with a focus towards U(1)-symmetric systems such as superconductors or ultra-cold atom gases in optical lattices~\cite{Podolsky11,Podolsky12,Pollet12,Chen13,Gazit14,Rancon14,Pekker15}. It has been concluded from analyzing both microscopic models and order-parameter field theories that the amplitude mode will be strongly masked, due to damping, in the order-parameter correlation function. However, a distinct signal of the amplitude mode can be isolated from accessing the so-called scalar susceptibility, in terms of correlations of the squared order-parameter field. For a magnetic systems, such as the one under consideration here, this implies that the amplitude mode, while not directly visible in the dynamical spin structure factor, should be accessible via appropriate scalar correlation functions, e.g., of bond-based spin-singlet operators~\cite{Podolsky11,Weidinger15}. The feasibility of using probes coupling to singlet observables in order to detect the amplitude mode in (three-dimensional) magnets, e.g. via light scattering, has recently been considered also within bond-operator mean-field theory~\cite{Matsumoto08,Matsumoto14}. From field-theoretical arguments it is expected that a marginally damped amplitude mode may be detected in close vicinity of the quantum critical point, while deep in the ordered regime it merges with the continuum of multi-magnon excitations, which exhibits a diverging spectrum (e.g. scaling as $\sim 1/ \omega$ with the frequency $\omega$ at the ordering wave vector~\cite{Zwerger04}), masking the amplitude mode.
While such general conclusions may be drawn based on field-theory considerations, it is of genuine interest -- also for future experimental probes -- to provide more quantitative microscopic details on the excitation spectra of two-dimensional quantum magnets across quantum critical points.

In this paper we use quantum Monte Carlo simulations combined with a stochastic analytic continuation approach to access the dynamical spin structure factor and adequate scalar response functions for the spin-half quantum Heisenberg model on the two-dimensional square-lattice bilayer. This model has been established~\cite{Singh88,Millis93,Millis94,Sandvik94,Chubukov95,Sommer01,Wang06} as a basic spin model exhibiting a quantum critical point: As a function of the ratio $g=J'/J$ between the interlayer ($J'$) to the intralayer ($J$) exchange coupling, it exhibits a quantum phase transition at a critical ratio~\cite{Wang06} $g_c=2.5220(1)$ that separates the small-$g$ collinear antiferromagnet from the large-$g$ quantum disordered dimer paramagnet.
This model has been analysed rather intensively in the past, and in particular large-scale quantum Monte Carlo simulations have identified both the above quoted location of the quantum critical point and verified the three-dimensional O(3) universality class of the quantum phase transition. In fact, due to the absence of geometric frustration, this system can be studied using quantum Monte Carlo without any sign-problem, even in close vicinity of the quantum critical point, by using by-now standard cluster update algorithms.
More recently, using an extended ensemble approach, also the quantum entanglement properties across the quantum phase transition have been analysed and possible universal contributions in the bipartite entanglement have been identified~\cite{Helmes14}.

However, no detailed account on dynamical properties based on quantum Monte Carlo simulations has been provided thus far. Clearly, detecting the characteristic excitation modes of both phases in realistic spectral probes is of most interest here. To access such dynamical properties based on quantum Monte Carlo simulations, one requires high-quality statistical data in order to perform the analytic continuation from the imaginary-time quantum Monte Carlo data to real frequencies.
We note that dispersion relations of the excitation modes have been studied previously, based on various analytical approaches such as  bond-operator mean-field theory or series expansions~\cite{Sommer01,Hamer12}. However,  a more direct relation to experimental probes requires also a quantitative evaluation of the spectral-weight distribution. In that respect, one of the fundamental questions is how one crosses over from two gapless spin-wave excitations in the limit of weakly coupled planes to gapped triplon excitations at strong interplane coupling.
This issue is of experimental relevance as well. For instance, the material BaCuSi$_2$O$_6$ is believed to realize\cite{sasago,bacusio,sebastian} the spin-half bilayer Heisenberg model on the square lattice  (interactions between the bilayers are weak and frustrated). These results are also expected to be of interest in the discussion of the bilayer iridate Sr$_3$Ir$_2$O$_7$\cite{kim,moretti}.

In the following, we provide a detailed account on the spectral-weight distribution in the dynamical spin structure factor and also address the detection of the amplitude mode by appropriate scalar response functions. The paper is organised as follows: in Sec.~II, we introduce the model Hamiltonian and the used computational and  analytical methods. The dynamical spin structure factor of the bilayer model is then analysed in Sec.~III, while Sec.~IV concentrates on the detection of the amplitude mode.
The results of our analysis are summarised in the concluding Sec.~V. Several computational details are provided in the appendices.


\section{Model and Methods}

The spin-half bilayer Heisenberg model on the square lattice is described by the Hamiltonian
\beq
\label{hh}
 H= J' \sum_i  \vec{S}_{i1} \cdot \vec{S}_{i2} + J \sum_{\langle i,j \rangle} (\vec{S}_{i1}\cdot\vec{S}_{j1}+\vec{S}_{i2}\cdot\vec{S}_{j2}),
\eeq
where spins $\vec{S}_{i\mu}$ reside on the lower ($\mu=1$) and upper ($\mu=2$) layer within the $i$-th unit cell of a square lattice, where the lattice constant is set to $a=1$ in the following. Note that each unit cell  contains an interlayer (rung) bond of the bilayer lattice.  Here, $J$ ($J'$) denote the intralayer (interlayer) exchange coupling. The model exhibits, in addition to the internal SU(2) spin symmetry and the square-lattice space-group symmetry, also a layer inversion symmetry in the layer indices ($1$ and $2$). We account for this additional quantum number when assigning a third component to an originally two-dimensional momentum space vector, such that $\vec{k}=(k_x,k_y,k_z)^{\intercal}$, with $k_z=0$ or  $\pi$, denoting the symmetric and antisymmetric channel with respect to layer inversion, respectively. Correspondingly, in position space, each spin is also assigned a transverse position, with respect to its layer index, such that $\vec{r}_{i\mu}$ is a
three-component position vector, with the third component equal to $0$ ($1$), for $\mu=1$ ($\mu=2$), respectively.

Of particular interest for our analysis is the dynamical spin structure factor, which is defined with respect to the Heisenberg-picture time evolution of the spin operators as (with $N_s$ denoting the number of spins)
\beq
S_S(\omega,\vec{k})= \frac{1}{N_s}\int dt\sum_{i,j,\mu,\nu} e^{i(\omega t - \vec{k}\cdot(\vec{r}_{i\mu}-\vec{r}_{j\nu}))} \langle \vec{S}_{i\mu}(t) \cdot\vec{S}_{j\nu}(0) \rangle.
\eeq
We will also refer to the $k_z=0$ ($k_z=\pi$) cases as the symmetric or even (antisymmetric or odd) sector.
In the presence  of long-range antiferromagnetic order, one may also distinguish  the components  of $S_S(\omega,\vec{k})$ parallel and transverse  with respect to the order parameter direction, in which case $S_S(\omega,\vec{k})$ represents a rotational average probed by  the quantum Monte Carlo simulations.

In addition to the spin-spin correlations, we also  consider in the following interlayer singlet bond (or spin-exchange) terms,
\beq
B_i=\vec{S}_{i1} \cdot \vec{S}_{i2},
\eeq
which define a corresponding scalar response function in terms of the dynamical structure factor  (with $N$ denoting the number of interlayer bonds,  and $N_s=2N$)
\beq
S_B(\omega,\vec{k})= \frac{1}{N}\int dt\sum_{i,j} e^{i(\omega t - \vec{k}\cdot(\vec{r}_{i}-\vec{r}_{j}))} \langle B_i(t) B_{j}(0) \rangle,
\eeq
where here $\vec{k}$ and the $\vec{r}_i$ denote two-dimensional square lattice k-space and lattice position vectors (i.e. with a vanishing third component), since the singlet operators $B_i$ reside at positions $\vec{r}_i$ on the square lattice spanned by the interplanar ($J'$) rung bonds. Given its scalar nature, we refer to $S_B(\omega,\vec{k}) $ also as the dynamical  singlet structure factor.

After having introduced the model and the relevant observables for our study, we next give here an overview of the various methods that we used. Details on the calculations with these methods can be found in the various appendices, as indicated below.

\subsection{Quantum Monte Carlo approach}

For the quantum Monte Carlo calculations, we used the stochastic series expansion method with directed loop updates~\cite{Sandvik99,Syljuasen02,Alet05}.
In the  simulations, we considered finite systems with $N_s=2L^2$ lattice sites and periodic boundary conditions in both square lattice directions $\vec{x}=(1,0)^\intercal$, and $\vec{y}=(0,1)^\intercal$.
In order to access ground state properties, the inverse temperature $\beta$ must  be chosen sufficiently large. This typically requires $\beta J\geq 2 L$. In our simulations, we considered mainly finite systems with $L=20$ with $N_s=800$ sites at $\beta J= 50$, unless specified otherwise.

To access the dynamical spin structure factor
using the quantum Monte Carlo simulations, we  efficiently~\cite{Michel07} measure the imaginary-time displaced spin-spin correlation functions directly in Matsubara frequency representation, which is related to $S_S(\omega,\vec{k})$ via
\beq\label{kernel}
S_S(i\omega_n,\vec{k})=\int_0^\infty {d\omega}\: \frac{\omega}{\pi} \frac{(1-e^{-\beta\omega})}{\omega^2+\omega_n^2}\:  S_S(\omega,\vec{k}).
\eeq
Here,  $\omega_n= 2\pi n /\beta$ for $n=0,1,2,...$  are bosonic Matsubara frequencies. One typically needs values of $n$ up to 160 to access the leading $1/\omega_n^{2}$ asymptotic behaviour of  $S_S(i\omega_n,\vec{k})$. The numerical inversion of Eq.~(\ref{kernel}) to obtain $S_S(\omega,\vec{k})$ from the Matsubara frequency quantum Monte Carlo data $S_S(i\omega_n,\vec{k})$ was performed using the stochastic analytic continuation method in the formulation of Ref.~\onlinecite{Beach04}. While such numerical analytic continuation methods tend to broaden inherent spectral features, they can still resolve a low number of excitation poles also from a separate continuum spectral contribution, depending on the quality of the imaginary-time data.

In order to efficiently access the dynamical singlet structure factor $S_B(\omega,\vec{k})$  of the bond-bond correlations from the quantum Monte Carlo simulations, we
measured  the corresponding bond-bond correlation functions in imaginary-time, binned over finite-width imaginary-time windows~\cite{Michel07}. Using an appropriate kernel for the analytic continuation, we then relate $S_B(\omega,\vec{k})$ directly to this imaginary-time binned quantum Monte Carlo data, without the need for an intermediate unfolding of the bin-resolution correlation function.
Further details of this measurement setup are provided in
App.~\ref{App:measure}.

\subsection{Spin-wave theory}

Upon expressing the spin fluctuations about the classical magnetically ordered state using the standard Holstein-Primakoff representation~\cite{Yosida96}, one arrives in the harmonic approximation at the  linear spin-wave description of the magnetic excitations in the ordered phase. As discussed in Sec. III, the comparison between the spin-wave results and the dynamical structure factor as obtained from the quantum Monte Carlo simulation improves upon taking into account corrections beyond the harmonic approximation in the spin-wave expansion. This has been done here by considering the next-to-leading order in the $1/S$ expansion of the Hamiltonian, and by performing a mean-field decoupling of the bosonic interaction terms to renormalize the coupling constants. Details on both the linear and higher-order spin-wave theory calculations are presented in App.~\ref{App:SWT}.

\subsection{Perturbation theory in $1/g$}

In the quantum-paramagnetic phase, an efficient description of the spin dynamics can be obtained from perturbation theory in the intralayer coupling $J$, i.e., starting from the limit of decoupled dimers which form a product state of singlets in the limit $J=0$. Exciting a single rung gives rise to a triplet excitation, which can delocalize for finite $J$, giving rise to a gapped excitation mode.
We perform a systematic expansion for the dynamical spin structure factor in $1/g$ up to second order for the triplet dispersion and to linear order in the spectral weight. While for the antisymmetric channel closed explicit expressions are obtained, we need to perform a numerical diagonalization of the effective interaction Hamiltonian in the two-triplon sector to access the
symmetric channel of the dynamical spin structure factor. Details on these calculations are provided in App.~\ref{App:PT} .

\subsection{Bond-operator-based $1/d$ expansion}

Bond operators~\cite{Sachdev90} were initially introduced to efficiently describe the paramagnetic phase of spin-half coupled-dimer magnets, the square lattice bilayer considered here being a prominent example. In this approach one introduces bosonic operators corresponding to spin-$1$ excitations (often called {\itshape triplons}) atop a singlet background~\cite{Kotov98, Collins06}. Generalisations of the bond-operator approach to magnetically ordered states~\cite{Sommer01} and to arbitrary spin~\cite{Ganesh11} have been investigated in the past. Moreover, it was recently shown~\cite{Joshi15a, Joshi15b} that bond operators enable a controlled description of coupled-dimer systems across the entire phase diagram using $1/d$ as a small parameter, where $d$ is the spatial dimension: relevant observables can be obtained in a systematic $1/d$ expansion once the dimer lattice has been generalized to $d$ space dimensions (for details we refer to Refs. ~\onlinecite{Joshi15a, Joshi15b}). In particular, the dispersion relation of triplon excitations in the disordered phase as well as that of gapless transverse modes and the gapped longitudinal mode in the ordered phase were reported along with their corresponding spectral weights in the dynamic spin susceptibility.
As will be discussed in Sec. III, a comparison with the dispersions obtained from quantum Monte Carlo  shows very good agreement after a {proper} mapping of the model parameters (the coupling ratio $g$), as explained in Sec.~III. In order to study the amplitude (Higgs) mode, we also calculated the bond-bond correlation  to leading order in $1/d$. A comparison with quantum Monte Carlo results will be discussed in Sec. IV. We find that to leading order in $1/d$,  the longitudinal mode gives rise to a single mode spectral contribution to the inter-layer singlet dynamical structure factor. Further details on the bond-operator calculations can be found in App.~\ref{App:BOT}.

\begin{figure*}[t]
\includegraphics[width=2\columnwidth]{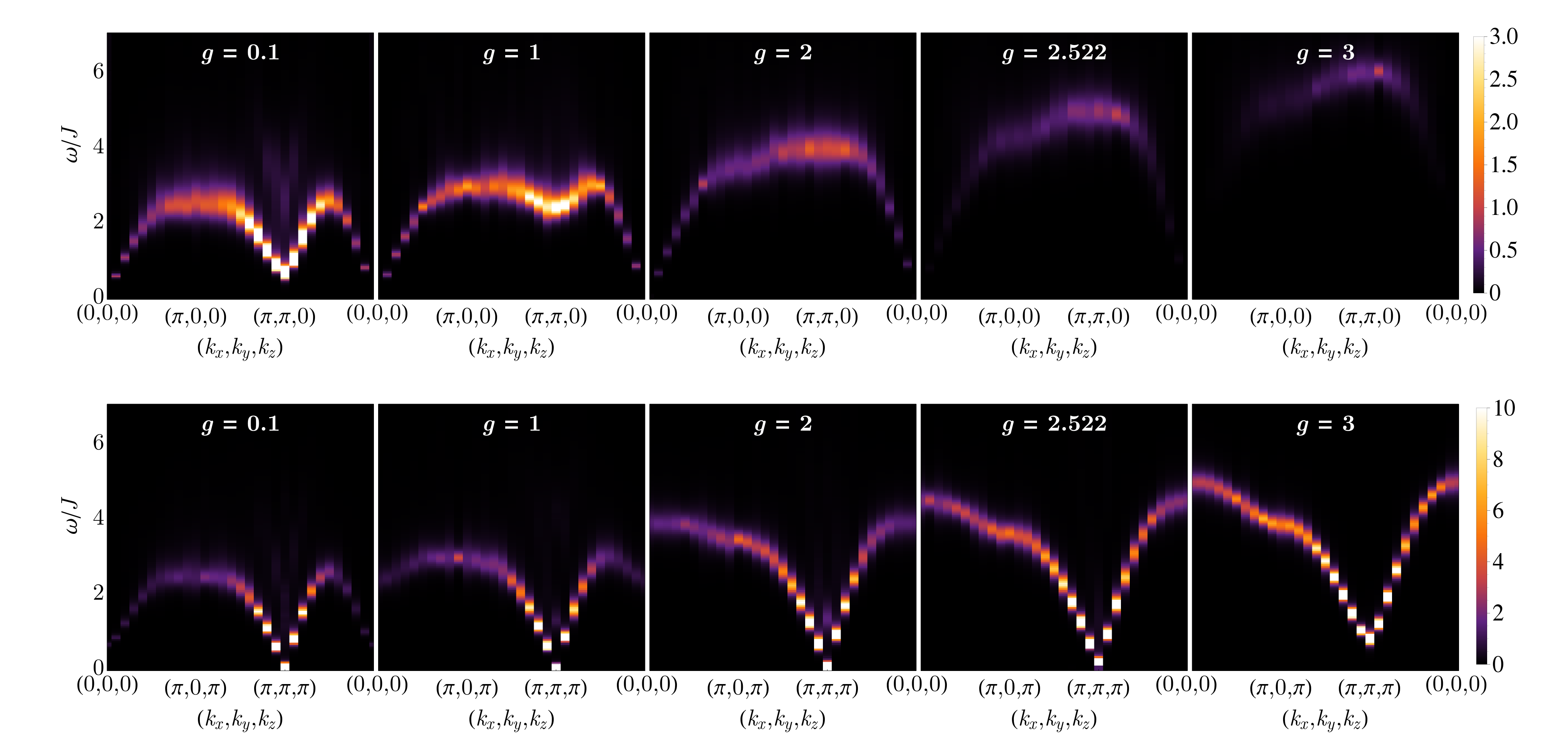}
\caption{Dynamical spin structure factor $S_S(\omega,\vec{k})$ in the symmetric (upper panel) and antisymmetric (lower panel) channel for the spin-half Heisenberg model on the square lattice bilayer at different ratios $g=J'/J$ of the interlayer ($J'$) to intralayer ($J$) coupling strength along the indicated path in the two-dimensional Brillouin zone.}
\label{fig:SS}
\end{figure*}


\section{Dynamical spin structure factor}\label{sec:SS}

The dynamical spin structure factor $S_S(\omega,\vec{k})$ is directly related to the scattering  intensity as probed by inelastic neutron scattering experiments.
Here,  we  monitor the evolution of  $S_S(\omega,\vec{k})$ upon varying the interaction ratio $g$ across the quantum critical point.

\subsection{Quantum Monte Carlo results}\label{sec:SSQMC}

We first present the results from the quantum Monte Carlo simulations for the dynamical spin structure factor.
In Fig.~\ref{fig:SS}, the dynamical spin structure factor $S_S(\omega,\vec{k})$ is shown for different values of $g$, each along a standard-path in the two-dimensional Brillouin zone, for the symmetric ($k_z=0$) and the antisymmetric ($k_z=\pi$) channel, separately.  The structure factor in the antisymmetric channel is dominated by a sharp single-mode-like contribution that softens in the antiferromagnetic regime at $\vec{k}=\vec{Q}$, the magnetic Bragg peak position
$\vec{Q}=(\pi,\pi,\pi)^\intercal$, while it exhibits a fully gapped branch in the quantum disordered regime, with a minimum gap $\Delta_T(g)$ in the dispersion, located at $\vec{k}=\vec{Q}$.
In the antiferromagnetic region, the dispersive feature traces the dispersion relation of the antiferromagnetic magnon (Goldstone) excitation-mode, while in the quantum disordered region, it follows  the gapped triplon dispersion. Both these relations will be substantiated by a direct comparison to spin-wave theory and perturbation theory calculations in $1/g$ for the antiferromagnetically ordered and quantum disordered regime respectively, as discussed below in Sec.~\ref{sec:SScompare}.

\begin{figure}[t]
\includegraphics[width=\columnwidth]{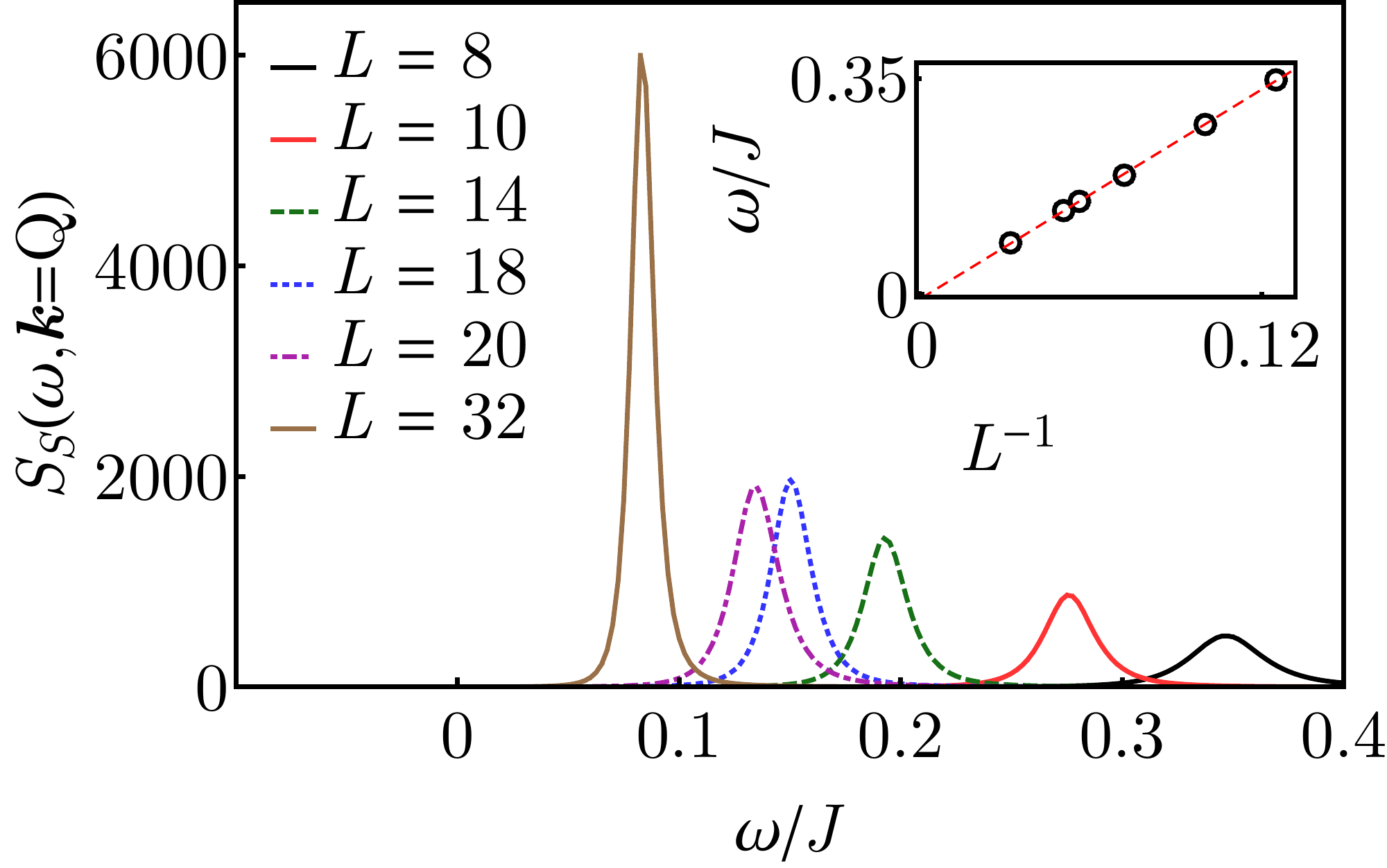}
\caption{(Color online) Dynamical spin structure factors  $S_S(\omega, \vec{k}=\vec{Q})$ for the spin-half Heisenberg model on the square lattice bilayer at the quantum critical point, $g=g_c$. The finite-size scaling of the excitation gap is  shown vs. $1/L$ in the  inset.}
\label{fig:SSscaling}
\end{figure}

Approaching the quantum critical point, we observe enhanced finite-size effects due to the algebraic growth of the  correlation length within the quantum critical region. We anticipate the most pronounced finite-size corrections  to appear right at the quantum critical point: based on the relativistic invariance  with a dynamical critical exponent $z=1$,  at low energies finite-size corrections of the peak position proportional to $1/L$ dominate at criticality. Indeed,  enhanced finite-size corrections at the quantum critical point were  reported recently for the low-energy dispersion of the Goldstone mode~\cite{Sen15}, and are visible in  Fig.~\ref{fig:SS} for $g=2.522$, where a small finite-size gap at $\vec{k}=\vec{Q}$ can be clearly resolved.
A  finite-size study of the dynamical spin structure factor, shown in  Fig.~\ref{fig:SSscaling}, indeed reveals a linearly vanishing finite-size excitation gap  as a function of $1/L$, as shown in the inset of that figure.

In addition to the sharp, single-mode feature,
the structure factor  also exhibits a broader distribution of spectral weight for energies above the single-mode threshold. This contribution will be discussed in more detail  in Sec. VI, since it is of  particular interest with regards to the possibility of observing a Higgs peak in the  spin structure factor.

The symmetric channel exhibits in the antiferromagnetically ordered phase a similarly sharply resolved distribution of the spectral weight, which however softens near the $\Gamma=(0,0,0)^\intercal$ point. Its dispersion follows to a very good accuracy that of the antisymmetric mode with a shift of the in-plane momentum by $(\pi,\pi)$. At the $\Gamma$ point, the spectral weight is completely suppressed for all values of $g$: this relates to the fact that spin fluctuations at the $\Gamma$ point, i.e. in the uniform magnetisation, vanish, due to the SU(2) symmetry of $H$. The gap at $\vec{k}=(\pi,\pi,0)^\intercal$ decreases upon approaching the limit of decoupled layers ($g\rightarrow 0$), and in the limit of decoupled layers, the structure factor in the symmetric channel softens at $\vec{k}=(\pi,\pi,0)^\intercal$, which is the single layer Bragg peak position.
By contrast to the antisymmetric channel, the overall spectral weight in the symmetric channel is strongly suppressed for large values of $g$: well in the quantum disordered regime, only a faint spectral weight distribution is present in the symmetric channel, while the antisymmetric channel provides a direct trace of the dispersion relation of the gapped triplon excitation mode.

The  bilayer Heisenberg model is obtained as the effective low-energy theory for the spin dynamics within the Mott insulator, in the strong-coupling limit of the half-filled Hubbard model on the square lattice bilayer. At strong local Hubbard repulsion $U\gg t$, we  thus expect that the evolution of the spectral weight in the dynamical spin structure factor, as described above, can be observed also in the Hubbard model, upon tuning the ratio $t'/t$ of the interlayer tunneling ($t'$) to the intralayer tunneling ($t$). It is less clear, however, that this also holds in the intermediate coupling regime, where $U$ is of order the bandwidth, and where residual charge fluctuations allow for higher-order spin exchange processes. To address this question, we performed also determinantal quantum Monte Carlo simulations for the Hubbard model on the square lattice bilayer. From our simulations, we find that also in the intermediate coupling region the dynamical spin structure factor exhibits the characteristic behaviour
that we observed in the Heisenberg limit. We provide details of these results for the Hubbard model in the supplemental material to this paper~\cite{suppmat}.
Returning here to the Heisenberg model, we next perform a detailed comparison of the quantum Monte Carlo results to the various theoretical approaches that we listed in Sec. II.

\subsection{Comparison to analytic results}\label{sec:SScompare}

In  the following, we  compare our quantum Monte Carlo data to the results obtained within (i)  linear and higher order spin-wave theory, (ii)  the  $1/g$ perturbation theory as well as (iii)  the $1/d$ expansion within the bond-operator approach.
While the spin-wave and the large-$g$ approach are restricted to the ordered and quantum disordered regimes, respectively, the $1/d$ expansion allows us to calculate appropriate structure factors throughout the whole phase diagram.
%
\begin{figure}[t]
\includegraphics[width=\columnwidth]{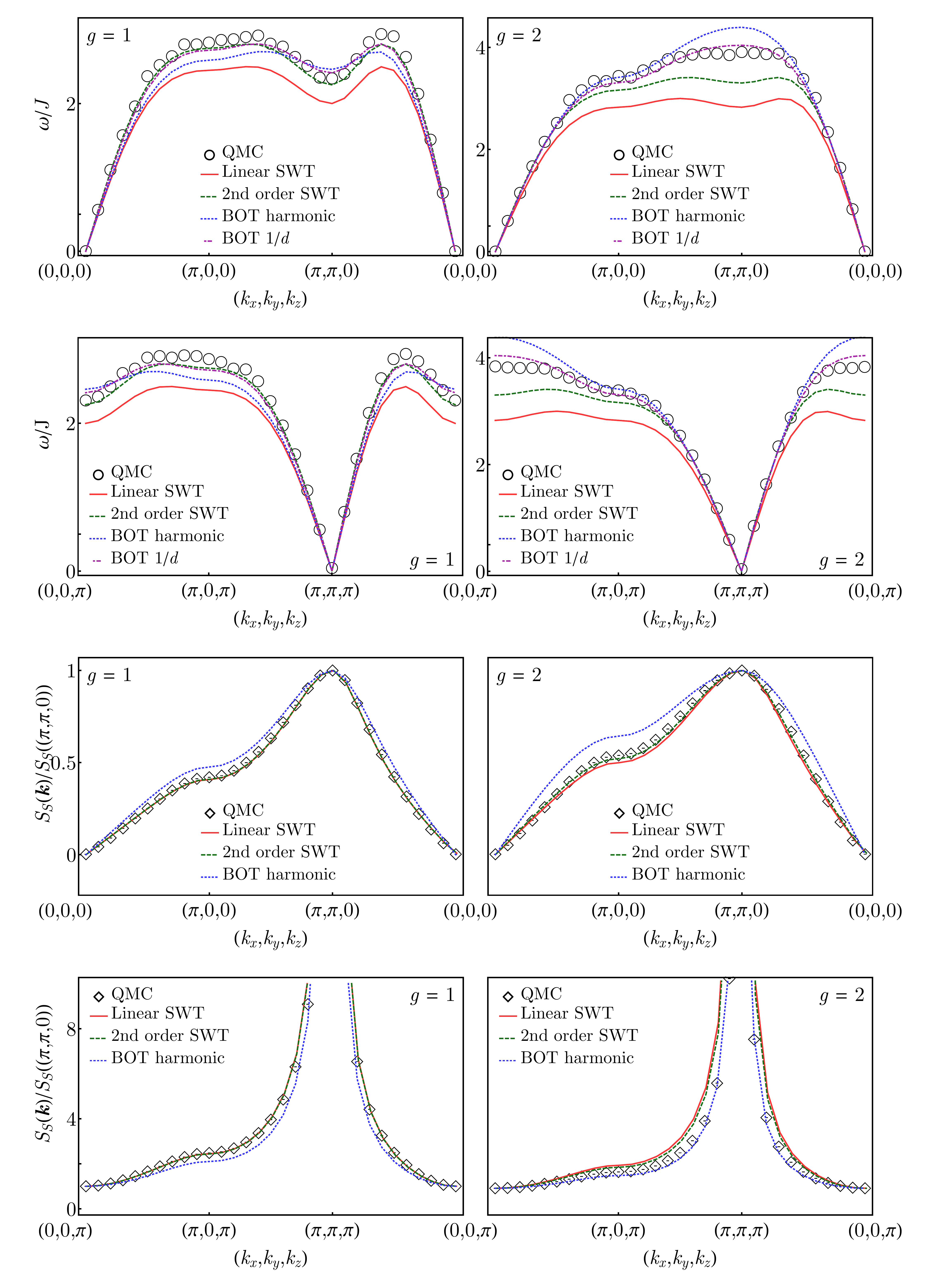}
\caption{(Color online) Comparison of the dynamical spin structure factor at different ratios $g=J'/J$ inside the antiferromagnetically ordered region for the spin-half Heisenberg model on the square lattice bilayer between the quantum Monte Carlo (QMC)  results and linear and 2nd order spin-wave theory (SWT) as well as bond-operator theory (BOT), both in harmonic approximation and including  leading $1/d$ corrections.  The upper panel compares the dispersion of the single-mode contribution, and the lower panel the integrated spectral weight $S_S(\vec{k})$. Here, the results for the symmetric (antisymmetric) channel have been normalised by the value of $S_S((\pi,\pi,0))$ ($S_S(0)$), corresponding to the maximum (minimum) value in that channel.
}
\label{fig:SScomp1}
\end{figure}
%
\begin{figure}[t]
\includegraphics[width=\columnwidth]{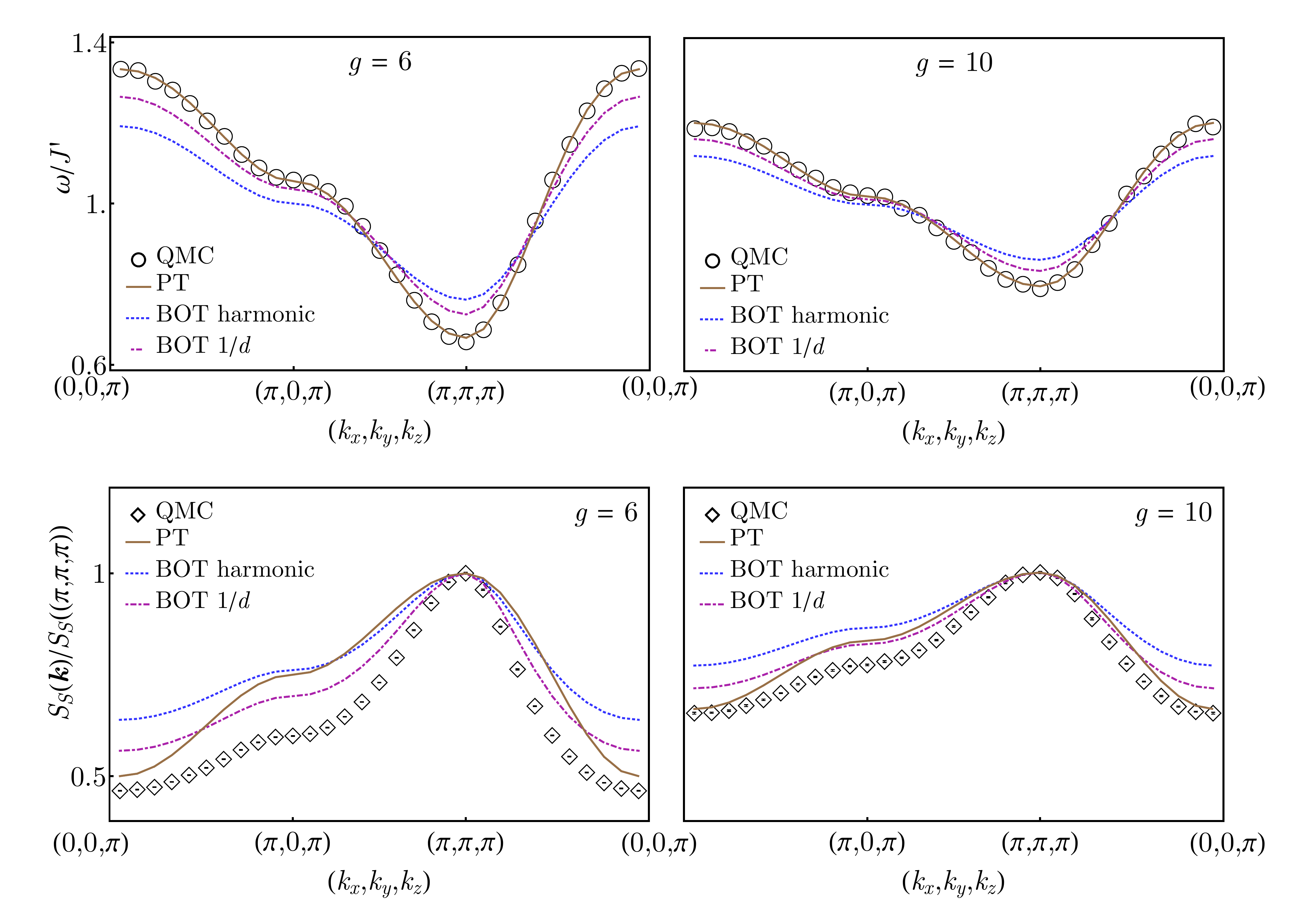}
\caption{(Color online) Comparison of the  dynamical spin structure factor in the antisymmetric channel at different ratios $g=J'/J$ well inside the quantum disordered  region for the spin-half Heisenberg model on the square lattice bilayer between the quantum Monte Carlo (QMC) results and $1/g$ perturbation theory (PT) as well as bond-operator theory (BOT) in both harmonic approximation and with leading $1/d$ corrections included. The upper panel compares the dispersion of the single triplon contribution, and the lower panel the integrated spectral weight $S_S(\vec{k})$. }
\label{fig:SScomp2}
\end{figure}
%
The upper two rows of Fig.~\ref{fig:SScomp1} and the upper row of Fig.~\ref{fig:SScomp2} show the dispersion relations for the region below and above the quantum critical point, respectively. In these panels, the mode dispersions from the quantum Monte Carlo data have been obtained by the peak positions in the spectral weight distribution, with the symbol size indicating the estimated uncertainty.
In addition to comparing the dispersions, we also compare, in the lower rows of  Fig.~\ref{fig:SScomp1} and Fig.~\ref{fig:SScomp2}, the integrated spectral weight
\beq
S_S(\vec{k})=\int d\omega \: S_S(\omega, \vec{k}),
\eeq
which in the quantum Monte Carlo simulations is conveniently obtained from the equal-time spin-spin correlations, to the results from spin-wave theory and the $1/g$ perturbation theory as well as the bond-operator theory. For the ordered phase, we performed the bond-operator theory calculations for the integrated spectral weight only in the  harmonic approximation, while for the disordered phase, we also considered the leading $1/d$ corrections. We also note that, in contrast to the mode dispersion, the integrated spectral weight can be obtained directly from the quantum Monte Carlo simulations without the need of performing the analytic continuation.

We first compare our numerical results to the dispersions obtained from spin-wave theory.
From the upper two panels of Fig.~\ref{fig:SScomp1}, we find that  (i) linear spin-wave theory, while providing a good overall account of the spin-wave dispersion, systematically underestimates the spin-wave energies, and (ii)  the higher order spin-wave theory approximation provides a significant improvement to the overall dispersions. As detailed in App.~\ref{App:SWT},
the net effect of the second order corrections to  linear spin-wave theory can be expressed in terms of  a $g$-dependent renormalization of the coupling ratio $g$, which for $g>1$ leads to an enhanced effective coupling ratio $g_\mathrm{eff}>g$, which results in the hardening of the spin-wave dispersion, as observed in Fig.~\ref{fig:SScomp1}. As discussed in App.~\ref{App:SWT}, we find that an even further, heuristic renormalization of
$g_\mathrm{eff}$ allows us to obtain even better matches of  the effective spin-wave theory dispersion to the numerical results.

With respect to the integrated spectral weight $S_S(\vec{k})$, which is shown in the lower two panels of
Fig.~\ref{fig:SScomp1}, we find that up to intermediate coupling ratios (cf. the case $g=1$ in the left panels  of Fig.~\ref{fig:SScomp1}), both linear and second order spin-wave theory provide a good overall account of the spectral weight distribution in the symmetric channel. Upon closer inspection however, one notices that the spin-wave theory results for $g=1$ fall slightly below the quantum Monte Carlo data for $S_S(\vec{k})$ for $\vec{k}$ near $(\pi,0,0)^\intercal$. This difference gets more pronounced upon increasing $g$ towards the quantum critical point, cf. the data for $g=2$ in the right panels of Fig.~\ref{fig:SScomp1}. The failure of low-order spin-wave theory to fully account for the spectral properties for
wave vectors near $\vec{k}=(\pi,0)^\intercal$ has been noticed previously for the case of a single layer Heisenberg model (i.e., $g=0$; cf., e.g., Ref.~\onlinecite{Sandvik01} for an extended discussion) and was recently linked to strong attractive magnon-magnon interactions that induce a Higgs resonance from two-magnon states~\cite{Powalski15}. One would then expect that
the increase in the deviation to low-order spin-wave theory that we observe in the integrated spectral weight for $g=2$ relates to
an enhancement in the Higgs resonance formation. If fact, as discussed in Sec. III, we can identify a well defined low-energy Higgs peak in the singlet structure factor for $g$ near $g_c$, which adds further support to this scenario.

For the antisymmetric mode, we find that both linear and higher-order  spin-wave theory accounts well for the integrated spectral weight distribution for $g=1$, while they overestimate the integrated spectral weight near the ordering wave vector,  $\vec{k}=\vec{Q}$, closer to $g_c$, cf. the data for $g=2$ in Fig.~\ref{fig:SScomp1}, reflecting the fact that quantum fluctuations of the order parameter are underestimated within these approximations.

Next, we focus on the quantum disordered region, and discuss the comparison to the $1/g$ perturbation theory.
Well inside  the quantum disordered region, we find that the $1/g$ perturbation theory, presented in App.~\ref{App:PT}, provides a rather good account of the quantum Monte Carlo data. A corresponding comparison of the triplon dispersion and the corresponding spectral weight (i.e. in the antisymmetric channel) for $g=6$ and $g=10$  is shown in Fig.~\ref{fig:SScomp2}. While the triplon dispersion is obtained very accurately within our $1/g$ perturbation calculations for both values of $g$, the integrated spectral weight at $g=6$ already exhibits somewhat larger quantitative differences than at $g=10$. As detailed in App.~\ref{App:PT}, we performed the perturbative calculations for the triplon dispersion up to quadratic order in $1/g$, while the spectral weight was obtained up to linear order, and thus is expected to be less accurate at larger values of $1/g$.
Nevertheless, the overall shape of the spectral weight distribution is well represented at this leading order already.

\begin{figure}[t]
\includegraphics[width=\columnwidth]{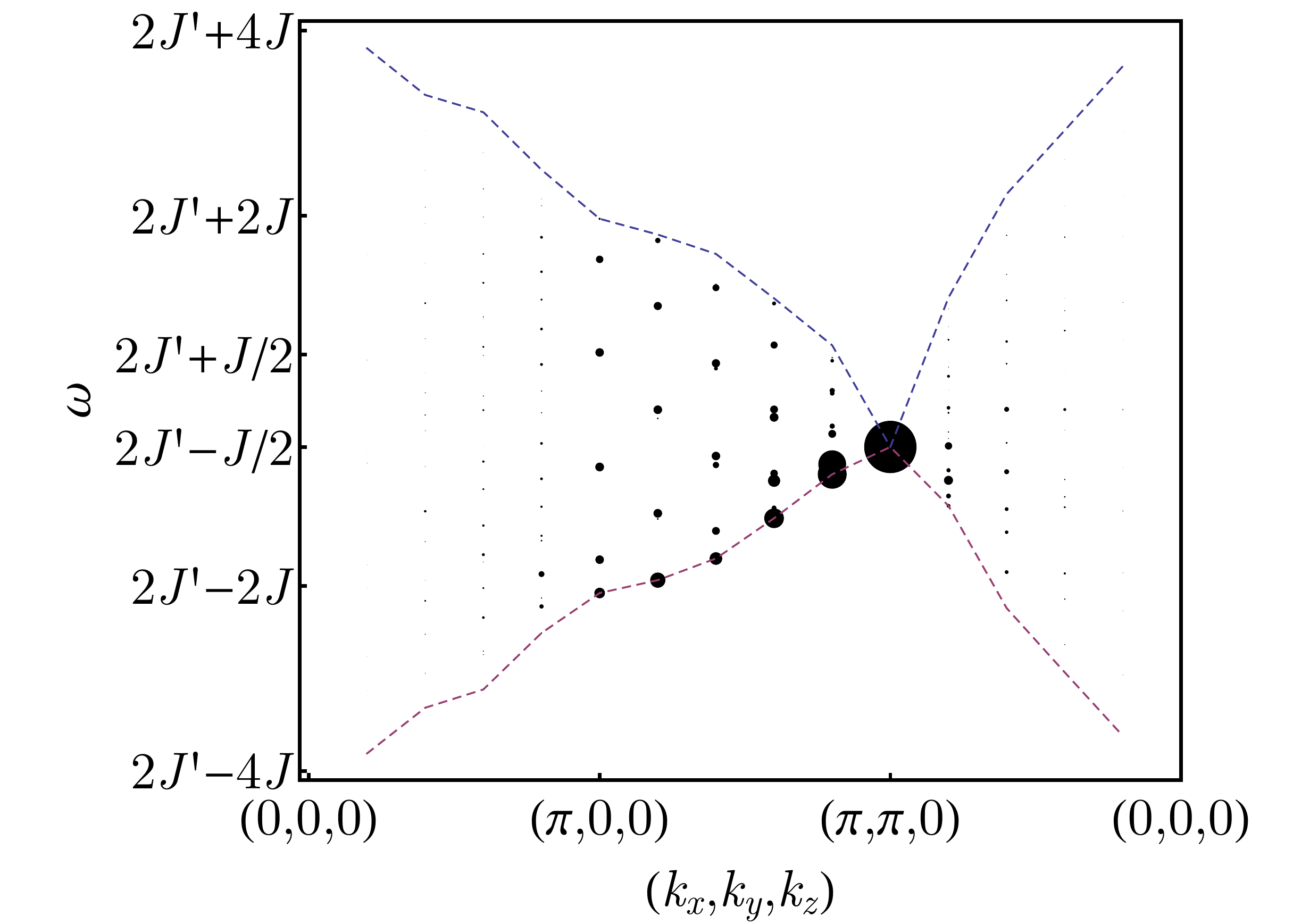}
\caption{(Color online) Distribution of the amplitude in the dynamical spin structure factor $S_S(\omega, \vec{k})$ for the symmetric sector ($k_z=0$) as obtained within the  $1/g$ perturbation theory for the numerical diagonalisation of the effective Hamiltonian within the two-triplon sector on a $L=10$ system.  The sizes of  circles are proportional to the amplitude at the centre positions. Dashed lines indicate the  upper and lower bounds of the continuum.}
\label{fig:SSsympt}
\end{figure}
Thus far, in discussing the quantum disordered region, we considered only the antisymmetric channel, which is dominated by the gapped triplon mode with a pronounced spectral weight.
However, within  $1/g$ perturbation theory, we  can also obtain a quantitative description of the faint spectral weight distribution for the symmetric channel in the quantum disordered regime.
Figure~\ref{fig:SSsympt} shows the result  for the dynamical spin structure factor  in the symmetric channel as obtained from  the $1/g$ perturbation expansion by a numerical diagonalization of the effective Hamiltonian within the two-triplon sector for a $L=10$ system. The structure factor is seen to display a concentration of the spectral weight at $\vec{k}=(\pi,\pi,0)^\intercal$, which corresponds to an eigenstate of the effective Hamiltonian in that sector. This is in good agreement with the quantum Monte Carlo data that also exhibit a concentration of the spectral weight for the symmetric channel near  $\vec{k}=(\pi,\pi,0)^\intercal$, cf., e.g., the data for $g=3$ in Fig.~\ref{fig:SS}.
One furthermore finds from the perturbative calculations,  that the amplitude of the symmetric channel is of order $(1/g)^2$, and thus vanishes in the limit $J/J'\rightarrow 0$ (cf. App.~\ref{App:PT} for details of the calculation), which is reflected also by the quantum Monte Carlo data.

The comparison of the quantum Monte Carlo data to the bond operator-based $1/d$ expansion requires an appropriate mapping from the coupling ratio $g$ to the parameter $q$ (defined as $q=J d/J'=d/g$, with $d$ the dimension of the system, here, $d=2$) that enters the $1/d$ expansion. Indeed, within the $1/d$ expansion, the quantum critical point is obtained as $q_c=1/2+3/(16d)+\mathcal{O}(1/d^2)$,  so that at the harmonic level (i.e. without  $1/d$ corrections)  $q_c=1/2$, while including first-order corrections in $1/d$ leads to $q_c=0.59375$,  to be compared to the  value of $(q_{QMC})_c=d/g_c=0.7930(2)$ that results from the (accurate) quantum Monte Carlo estimate of $g_c$. We need to account for this difference when performing the comparison, in particular in the vicinity of $g_c$.
For a given value of $g\leq g_c$, we fix the value of $q$ such that  $(q-q_c)$ equals the corresponding absolute distance in $q_{QMC}=d/g$ to $(q_{QMC})_c$, i.e.,
such that
\beq
q-q_c=q_{QMC}-(q_{QMC})_c .
\eeq
In the quantum disordered phase, $g>g_c$, it is more appropriate to relate the parameters by considering a fixed \emph{relative} distance to the quantum critical point, i.e., a value of $q$ is obtained, such that
\beq\label{eq:relativemapping}
\frac{q-q_c}{q_c}=\frac{q_{QMC}-(q_{QMC})_c}{(q_{QMC})_c}. 
\eeq
The corresponding mode dispersions are shown in the upper panels of Fig.~\ref{fig:SScomp1} for $g<g_c$ and Fig.~\ref{fig:SScomp2} for $g>g_c$, respectively, while the lower panels of these figures show the comparison in the integrated spectral weight $S_S(\vec{k})$. We find that with the above parameter mappings, the  bond-operator theory provides a good qualitative account of both the spin-wave dispersion and the triplon dispersion within this unified approach, at least when the leading
$1/d$ corrections beyond the harmonic approximation are taken into account. Nevertheless, the harmonic approximation already keeps track of the leading dispersive features and provides a good account of the integrated spectral weight in both phases.  For the following discussion of the amplitude mode, we thus limited the bond-operator theory calculations to the harmonic level.


\section{Amplitude (Higgs) mode}\label{sec:SB}

\begin{figure*}[t]
\includegraphics[width=2\columnwidth]{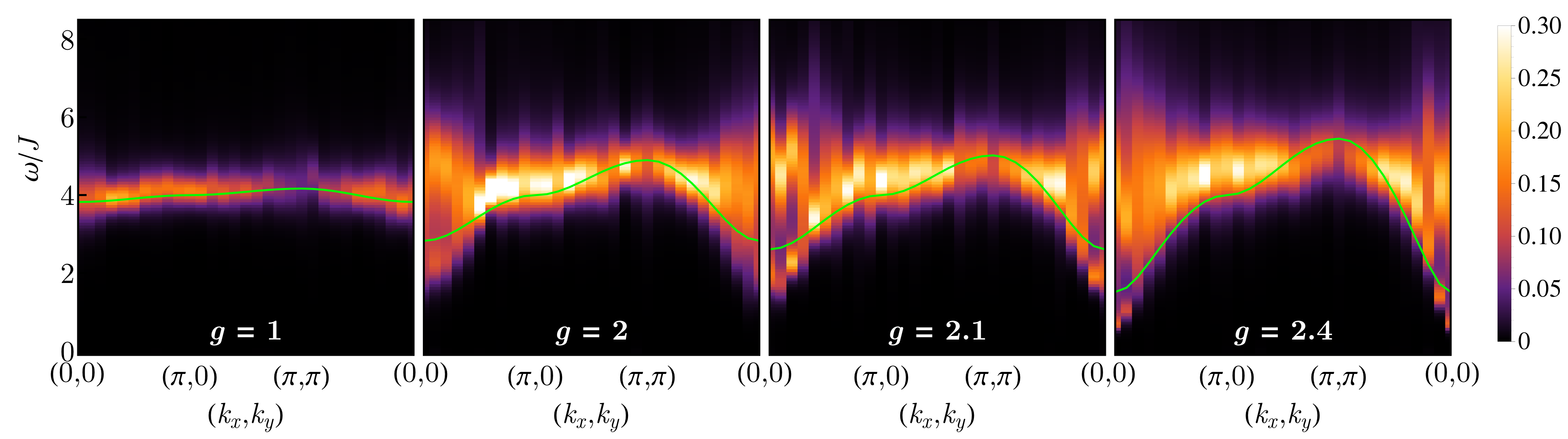}
\caption{(Color online) Dynamical  singlet structure factor $S_B(\omega,\vec{k})$  for the spin-half Heisenberg model on the square lattice bilayer at different ratios $g=J'/J$ of the interlayer ($J'$) to intralayer ($J$) coupling strength along the indicated path in the two-dimensional Brillouin zone. Also indicated by lines are the amplitude-mode dispersions obtain within the bond-operator theory.}
\label{fig:SBcomp}
\end{figure*}

The longitudinal, amplitude fluctuations of the order parameter field need to be properly taken into consideration in a quantitative  description  of the quantum critical bilayer Heisenberg antiferromagnet and its low energy properties~\cite{Chubukov95,Sommer01}. For example, within bond-operator theory calculations, the amplitude mode is found to soften upon approaching the quantum critical point, thus restoring the $SU(2)$ symmetry, while it hardens towards an elevated energy scale of order $4J$ in the limit of decoupled layers ($J'=0$), rendering it insignificant with respect to the order-parameter fluctuations well inside the ordered phase~\cite{Chubukov95,Sommer01}.
While the relevance of the amplitude mode on e.g. the order parameter strength and the critical value of $g$ has been well studied in the past, only recently have possible direct signatures of the amplitude mode and  appropriate experimental probes been addressed. In particular, it has been argued that response functions which couple to the square of the order parameter will show a distinct peak from the amplitude mode (Higgs peak),  well separated from the multi-magnon contribution to the overall spectral weight in the ordered phase, close to the quantum critical point~\cite{Podolsky11,Weidinger15}.

\subsection{Quantum Monte Carlo results}\label{sec:SBQMC}
To quantify these considerations for the bilayer Heisenberg model,
we examine in this section the interlayer-bond spin-exchange correlations, in terms of the corresponding dynamical singlet structure factor $S_B(\omega,\vec{k})$ introduced in Sec. II. Based on its scalar nature, we expect the correlations of this observable to exhibit in addition to possible multi-magnon contributions also a distinct signal from the amplitude mode, most pronounced in the low-energy regime, and within the vicinity of the quantum critical point.
In Fig.~\ref{fig:SBcomp}, we show our numerical data for different values of $g$, taken on a $L=20$ system at $\beta J=50$, representative of ground state expectation values.
We obtain for all values of $g$ a spectral contribution to $S_B(\omega,\vec{k})$ at elevated energies $\omega\approx 4J$, which does not exhibit a strong dependence on the coupling ratio $g$. In addition however, we observe in the vicinity of the quantum critical point a distinct feature in the low-energy region,  i.e., separated from the higher-energy signal. This additional spectral branch is most pronounced in the vicinity of the $\Gamma$ point, where also its spectral weight is most clearly visible.
\begin{figure*}[t]
\includegraphics[width=2\columnwidth]{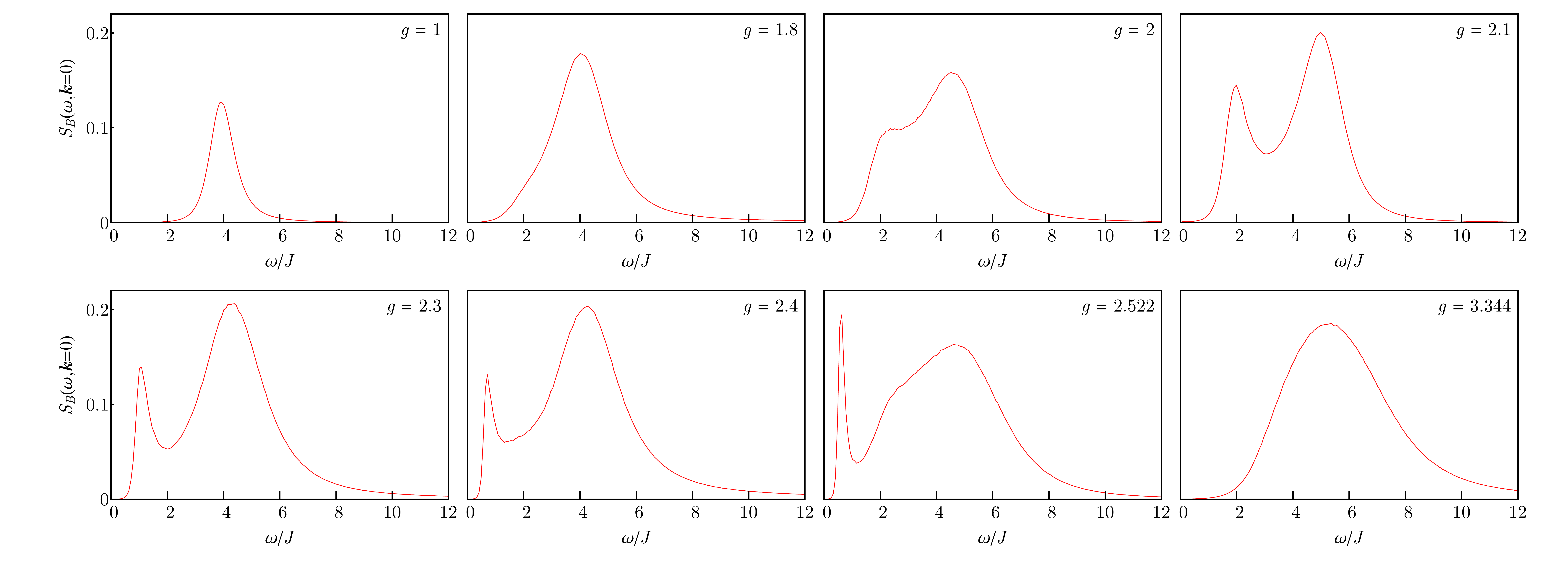}
\caption{(Color online) Dynamical  singlet structure factor $S_B(\omega,\vec{k}=0)$ at the $\Gamma$ point for the spin-half Heisenberg model on the square lattice bilayer at different ratios $g=J'/J$ of the interlayer ($J'$) to intralayer ($J$) coupling strength.}
\label{fig:SBGamma}
\end{figure*}

\begin{figure}[t]
\includegraphics[width=\columnwidth]{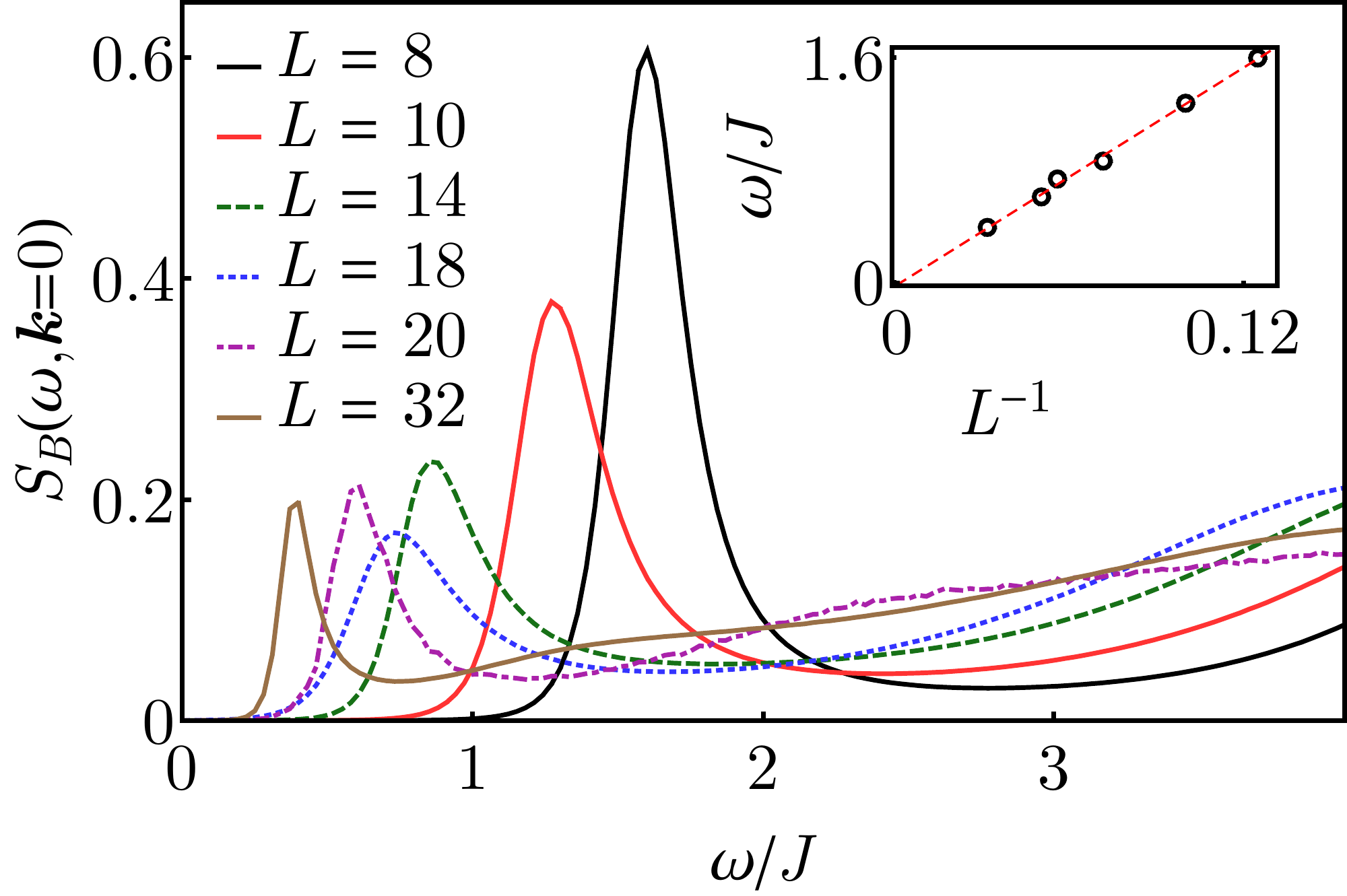}
\caption{(Color online) Dynamical  singlet structure factor $S_B(\omega, \vec{k}=\vec{Q})$ for the spin-half Heisenberg model on the square lattice bilayer at the quantum critical point, $g=g_c$. The finite-size scaling of the excitation gap is shown vs. $1/L$ in the  inset.}
\label{fig:SBscaling}
\end{figure}

To better exhibit the emergence of the low-energy Higgs peak from the multi-magnon contribution, we show in Fig.~\ref{fig:SBGamma} the dynamical singlet structure factor $S_B(\omega,\vec{k}=0)$ at the $\Gamma$ point for different values of the coupling ratio $g_c$. Within the range $2 \lesssim g \lesssim g_c$,
we detect a separate, low-energy peak split-off from the broad second peak in the spectral weight at elevated energies. The  position of the low-energy peak tends towards lower energies upon tuning $g$ towards its critical value $g_c$.
However, approaching  the quantum critical point, we again observe enhanced finite-size effects, most pronounced at the quantum critical point itself. This behavior is illustrated for $g=g_c$ in  Fig.~\ref{fig:SBscaling}. We find that
for $g$ below about $g \approx 2$, the Higgs peak merges with the second peak, so that a Higgs peak cannot be discerned anymore as a separate spectral feature, restricting its observation to the close vicinity of the quantum critical point.
%
%
\begin{figure}[t]
\includegraphics[width=\columnwidth]{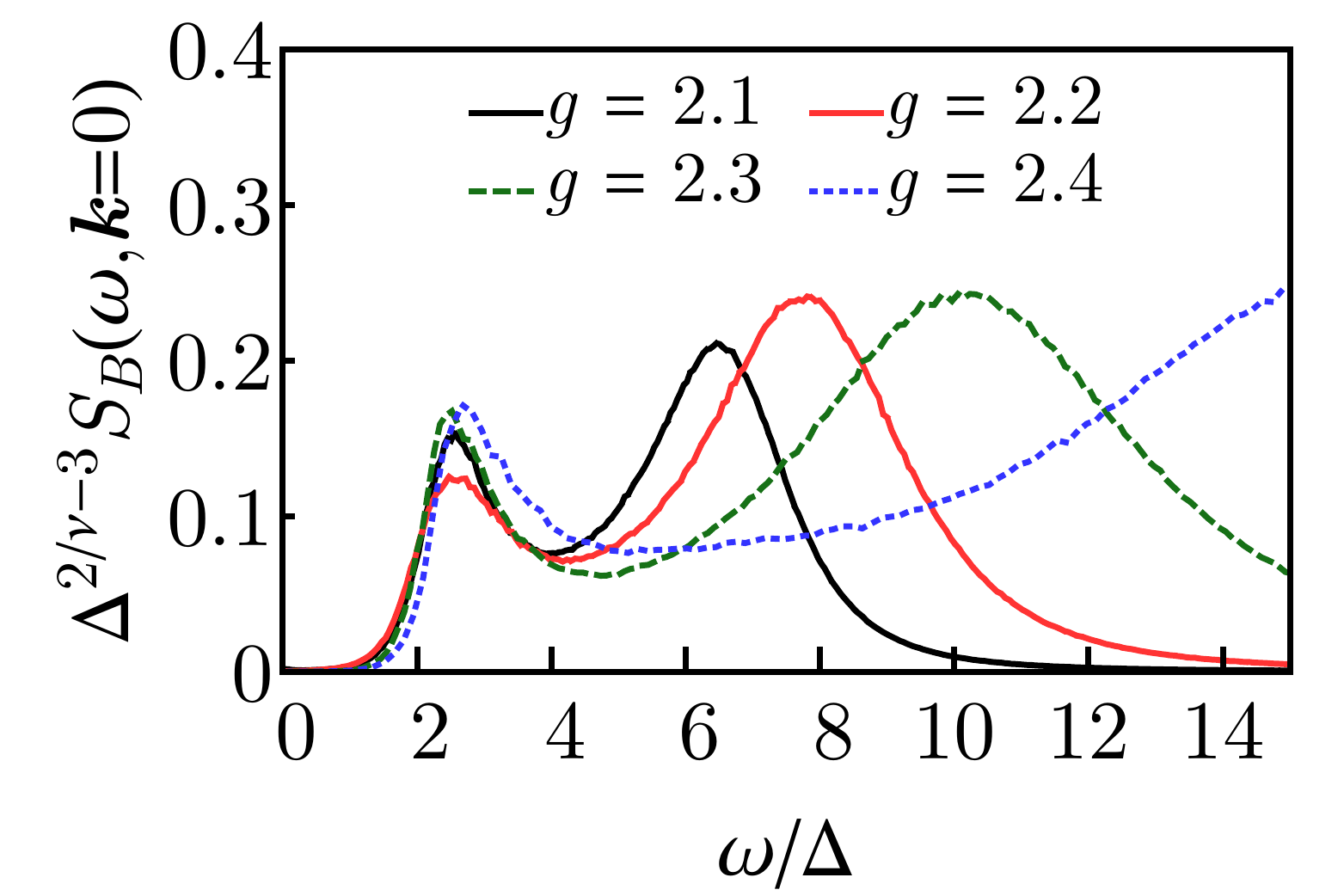}
\caption{(Color online) Scaling plot of the dynamical singlet structure factor $S_B(\omega,\vec{k}=0)$ at the $\Gamma$ point for the spin-half Heisenberg model on the square lattice bilayer at different ratios $g=J'/J$ of the interlayer ($J'$) to intralayer ($J$) coupling strength near the quantum critical pint $g_c$.}
\label{fig:SBscale}
\end{figure}
%

\subsection{Comparison to bond-operator theory}\label{sec:SBcomparison}

Within the parameter regime for which we can identify a Higgs-mode peak at the $\Gamma$ point, we also observe traces of the corresponding mode dispersion near $\Gamma$. It is thus interesting to compare our quantum Monte Carlo data to the amplitude-mode dispersion obtained from the bond-operator-based $1/d$-expansion~\cite{Joshi15b}. Referring for details on the $1/d$-expansion calculations of $S_B$ to App.~\ref{App:BOT}, we compare in Fig.~\ref{fig:SBcomp} the resulting dispersion relation for the amplitude mode with the quantum Monte Carlo data, using the parameter mapping from Eq.~(\ref{eq:relativemapping}).

Using bond-operator theory we also calculated the spectral weight of the amplitude mode within the harmonic approximation. As discussed in more detail in App.~\ref{App:BOT},  one finds an enhanced spectral weight in $S_B(\omega,\vec{k})$ near the $\Gamma$ point, in particular in the vicinity of  the quantum critical point, a result which appears to agree with the quantum Monte Carlo data, which also exhibit more pronounced spectral weight for the amplitude mode branch near  the $\Gamma$ point and near the quantum critical point. A more quantitative comparison is limited by the fact that there are contributions (of multi-mode character) to the spectral weight observed in $S_B$ beyond the pure Higgs-mode contribution considered in App.~\ref{App:BOT}.

\subsection{Scaling form and comparison to $S_S(\omega, \vec{k})$}

To analyse further the shape of the Higgs peak, we compare our numerical results to  the universal low-energy scaling form of the scalar response function~\cite{Podolsky12} obtained from a $1/N$-expansion of the  O(N) model  in the quantum critical region,
\beq\label{eq:scalingscalar}
S(\omega)\propto \Delta^{3-2/\nu}\Phi(\omega/\Delta),
\eeq
in terms of the characteristic energy scale in the quantum critical region, $\Delta\propto (1-g/g_c)^\nu$, with the correlation length exponent $\nu$ of the three-dimensional O(3) universality class, to which the quantum critical point $g_c$ in the bilayer Heisenberg model belongs. The value of the critical exponent
$\nu$ has been determined by previous large scale Monte Carlo simulations~\cite{Hasenbusch01} as $\nu=0.710(2)$.
Finally, $\Phi$ denotes a corresponding scaling function.
We estimate the $g$-dependent excitation gap of the amplitude mode $m_H$(g) (Higgs mass) from the  position of the low-energy peak in $S_B(\omega,\vec{k}=0)$, and the energy scale $\Delta(g)$ for $g<g_c$ in terms of the triplet excitation gap $\Delta_T(g')$ in the quantum disordered phase, where $g'>g_c$ is the mirrored (with respect to the quantum critical point) coupling ratio defined by $g'-g_c=g_c-g$. In other words, $\Delta(g)=\Delta_T(g_c+(g_c-g))$. For $2<g<g_c$, i.e. in the region where we can identify a Higgs peak, we find that $m_H(g)/\Delta(g)=2.6(4)$, in agreement with a previous estimate for this ratio from the effective O(3) field theory~\cite{Gazit14}.

\begin{figure*}[t]
\includegraphics[width=2\columnwidth]{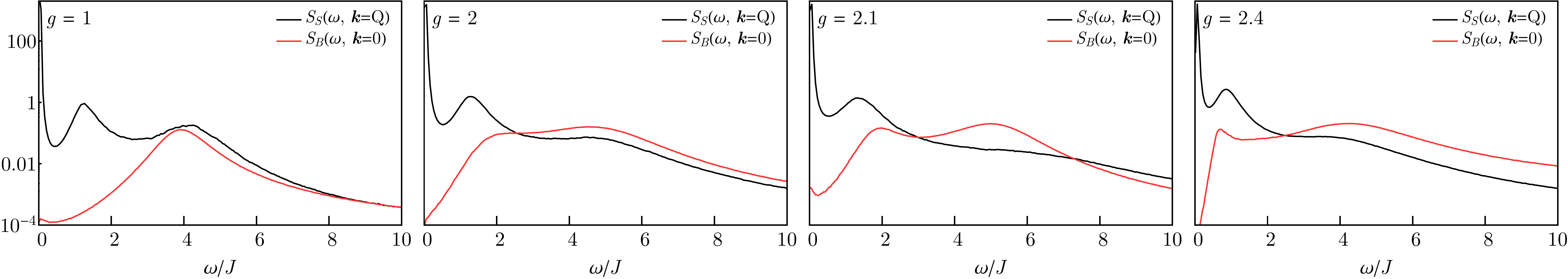}
\caption{(Color online) Comparison between the dynamical spin structure factor $S_S(\omega, \vec{k}=\vec{Q})$ and the dynamical singlet structure factor $S_B(\omega, \vec{k}=0)$ for the spin-half Heisenberg model on the square lattice bilayer at different ratios $g=J'/J$ of the interlayer ($J'$) to intralayer ($J$) coupling strength.}
\label{fig:SSSB}
\end{figure*}

Based on this relation, we show in Fig.~\ref{fig:SBscale} the appropriately rescaled data for $S_B(\omega,\vec{k}=0)$ from different values of $g$ according to the scaling law in Eq.~(\ref{eq:scalingscalar}). From this analysis we find that the low-energy  part of  $S_B(\omega,\vec{k}=0)$  relates well to this scaling form, adding further support to the interpretation of the low-energy peak in terms of the amplitude mode.
By contrast, the second peak at energies $\omega\approx 4J$ does not follow this scaling form, hence is not part of the  universal signal  in the response function. In fact, as already mentioned above, we see from Fig.~\ref{fig:SBGamma} that the position of the second peak does not change much upon varying $g$, in contrast to the Higgs peak.
The observation, that the amplitude mode follows the scalar response function scaling form also implies that in this two-dimensional system, the amplitude mode is only marginally damped in the vicinity of the quantum critical point, i.e. with a width that scales like the Higgs mass $m_H$ near the quantum critical point, thus making the observation of this excitation mode possible.

We find that the quantum Monte Carlo data are consistent also  with the low-energy scaling prediction~\cite{Podolsky12} $S(\omega)\propto \omega^3$ in as much as  the imaginary-time ($\tau$) data for the bond-bond correlations at $\vec{k}=0$ (which are actually calculated in the simulation) agrees with the corresponding decay  $S_B(\tau,\vec{k}=0) \propto 1/\tau^4$ at large imaginary times (cf.  App.~\ref{App:measure} for details). However, we cannot independently estimate the large-$\tau$ scaling behaviour since we are limited by enhanced statistical noise on the imaginary-time data in the relevant $\tau$ regime (cf. App.~\ref{App:measure}).

In order to contrast further the appearance of the amplitude mode in the  dynamical singlet structure factor as compared to the dynamical spin structure factor, we show for different values of $g$ in Fig.~\ref{fig:SSSB}  both (i) the dynamical singlet structure factor $S_B(\omega,\vec{k}=0)$ at the
wave vector where $S_B$ softens at $g_c$, and (ii) the dynamical spin structure factor
$S_S(\omega,\vec{k}=\vec{Q})$ at the Bragg peak position where a possible Higgs mode-contribution would soften at the quantum critical point~\cite{Joshi15b}.

As mentioned already in Sec.~\ref{sec:SSQMC}, in the ordered phase, $S_S(\omega,\vec{k})$ exhibits in addition to the magnetic Bragg peak an extended tail of spectral weight.
This is shown for $\vec{k}=\vec{Q}$ for several  values of $g$  in Fig.~\ref{fig:SSSB}.  Atop a background signal that falls off with increasing $\omega$, and which is expected from general considerations in the symmetry-broken phase~\cite{Zwerger04}, we observe two broad additional features in $S_S(\omega, \vec{Q})$ at $\omega\approx J$ and $\omega\approx 4J$, respectively. Compared to the magnetic Bragg peak, both these features are significantly broader and of suppressed spectral weight. These features do not exhibit systematic trends upon varying $g$, apart from weak shape variations for different values of $g$. By contrast, the data for $S_B(\omega,\vec{k}=0)$ exhibits a clear Higgs peak split-off at $g\approx 2$, that we already analysed above, with a $g$-dependent peak position and a shape that agrees with the scalar response function scaling form.
Fig.~\ref{fig:SSSB} illustrates how the presence of the  extended residual  spectral weight  tail beyond the magnetic Bragg peak  masks a possible observation of the low-energy Higgs peak in the dynamical spin structure factor even near criticality, in contrast to the scalar response function, which is strongly suppressed at low energies. Furthermore, we find that both structure factors show an enhanced spectral weight in the form of a broad peak at similarly elevated energies $\omega\approx 4J$, suggesting that the spectral weight observed in $S_B(\omega,\vec{k})$ in this energy range is dominated by multi-magnon contributions.

We  finally note that we also examined using the quantum Monte Carlo simulations the intraplanar bond-bond correlations, and find  in the symmetric sector (with respect to layer inversion) a similar  (but weaker) amplitude-mode contribution emerging in the  low-energy region, while no such feature is observed for the antisymmetric channel. As noted in App.~\ref{App:BOT}, this observation also agrees with the bond-operator-based $1/d$ expansion calculations of the intraplanar bond-bond correlations.


\section{Conclusions}

Using quantum Monte Carlo simulations and a variety of analytical approaches, we have obtained precise information on the dynamical
response functions of the spin-1/2 Heisenberg model on a bilayer system of square lattices. We have shown that the transition from two gapless
spin-wave excitations in the limit of weakly coupled planes to a single gapped triplon excitation in the strong coupling limit occurs via a rapid suppression
of the spectral weight of the symmetric spin-wave mode upon increasing the interlayer coupling. The intensity of that mode in the dynamical spin structure factor probed in neutron scattering is already extremely weak at the quantum critical point where the system becomes gapped, and it is
the antisymmetric mode that smoothly connects to the triplon mode of the strong coupling limit. We have also looked for traces of the amplitude (Higgs)
modes in various dynamical correlation functions. This mode is clearly visible as a distinct spectral feature in the interlayer-bond dynamical singlet structure factor,  which opens to the way to its detection with light scattering. However, this mode is not visible in the dynamical spin structure factor, most likely because
it is masked by an almost dispersionless continuum of incoherent excitations. It is our hope that these conclusions will further motivate the experimental
investigation of the dynamical response of bilayer systems.

\subsection*{Acknowledgments}
We thank  G. Jackeli, G. Khaliullin, A. L\"auchli, K. P. Schmidt, V. Schnells, and A. V\"oll for discussions
and acknowledge support by the Deutsche Forschungsgemeinschaft (DFG) under grant FOR 1807 (FFA, ML, SW) and SFB 1143 (DGJ, MV) and by the Swiss National Science Foundation. Furthermore, we thank the IT Center at RWTH Aachen University and the JSC J\"ulich for access to computing time through JARA-HPC and on JUROPA.
SW thanks the KITP Santa Barbara for hospitality during the program ``Entanglement in Strongly-Correlated Quantum Matter''. This research was supported in part by the National Science Foundation under Grant No. NSF PHY11-25915.


\appendix
\section{Bond operator measurements in quantum Monte Carlo}\label{App:measure}

\begin{figure}[t]
\includegraphics[width=0.8\columnwidth]{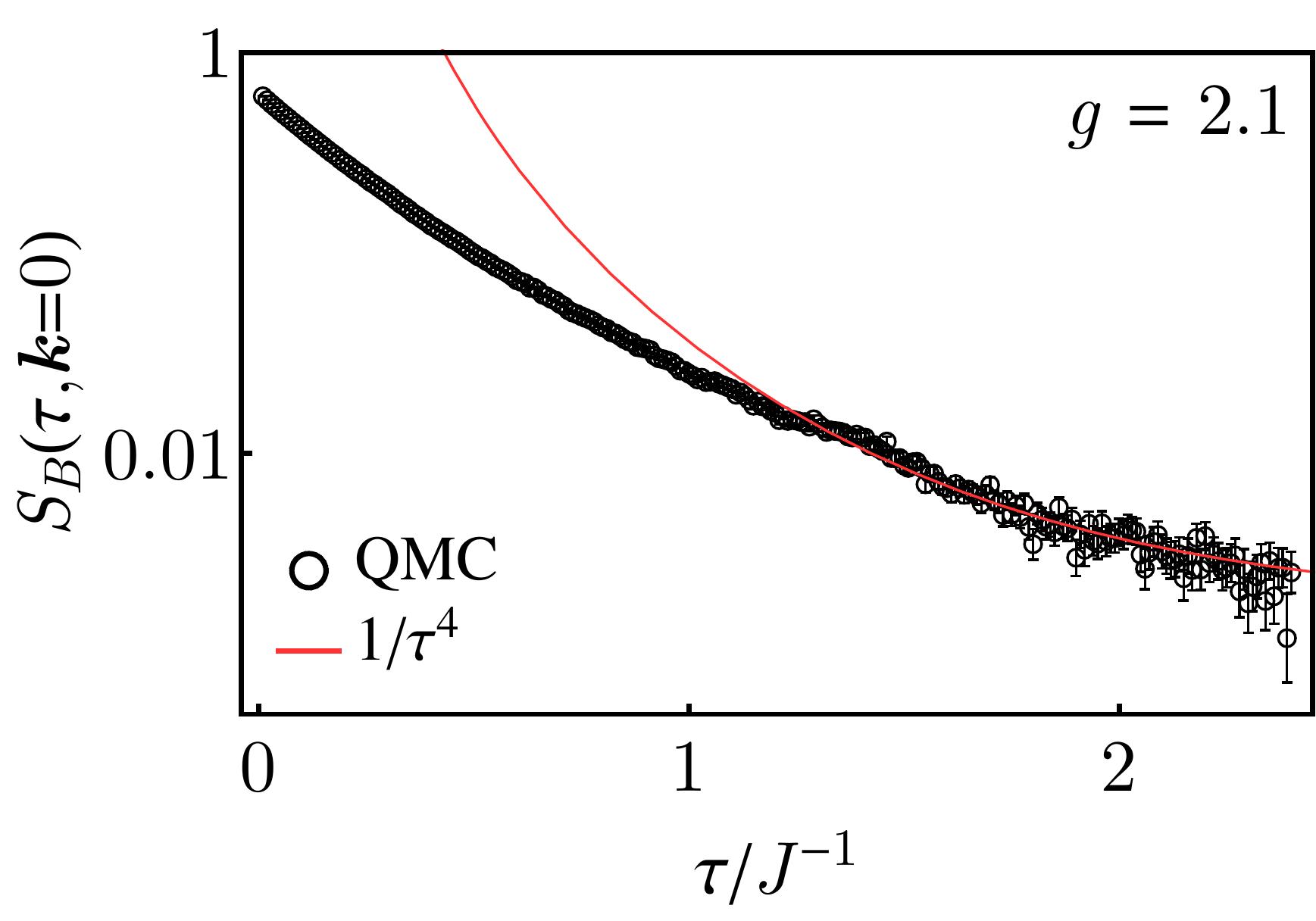}
\caption{(Color online) Imaginary-time ($\tau$) dependence of  $S_B(\tau,\vec{k})=\langle B_\vec{k}(\tau) B_{-\vec{k}} \rangle$ at the $\Gamma$ point, $\vec{k}=0$, for a coupling ratio of $g=2.1$, along with an asymptotic $1/\tau^4$-decay, indicated by the line.}
\label{fig:SBtau}
\end{figure}

To efficiently calculate the  dynamical singlet structure factor $S_B(\omega,\vec{k})$ within the stochastic series expansion (SSE) formulation of quantum Monte Carlo simulations, we extend an approach put forward in Ref.~\onlinecite{Michel07}. It is based on the well-known mapping of the discrete SSE configuration space onto a continuous-time  world-line formulation\cite{Sengupta02}. For hermitian conjugate operators $A$ and $B$ that are constructed from Hamiltonian terms, such as the Fourier modes
\beq
B_\vec{k}=\frac{1}{\sqrt{N}}\sum_i e^{-i\vec{k}\cdot\vec{r}_i} B_i
\eeq
of the spin-exchange terms of  the bond-operators $B_i$ introduced in Sec. II, instead of the imaginary-time ($\tau$) correlation function $\langle A(\tau) B \rangle$, we measure an auxiliary  correlation function that is defined in terms of a discrete time grid laid  over the   $\tau$-interval $[0,\beta]$, with width $\delta\tau$. Namely, we consider the Monte Carlo estimator
\beq
C_{AB}(l \: \delta\tau)=\left\langle\frac{1}{\delta\tau^2}\:N_{A \in [l\delta\tau,(l-1)\delta\tau]}\:N_{B \in [\delta\tau,0] }\right\rangle_\mathrm{MC},
\eeq
where $\langle\cdot\rangle_\mathrm{MC}$ denotes the  Monte Carlo expectation value, and $N_{A\in I}$ counts the number of operators $A$ in the SSE operator sequence which lie within the time imaginary-time interval $I$, while $l=1,...,L_\tau$, with $L_\tau$ given by the number of time-slices, $\beta=L_\tau\:\delta\tau$. Using the periodicity of the SSE configuration, one can further improve the above estimator to
\beq
C_{AB}(l \: \delta\tau)=\frac{1}{\beta\delta\tau}\left\langle N_{(A,B),l}\right\rangle_\textrm{MC},
\eeq
where $N_{(A,B),l}$ counts the number of pairs $(A,B)$ in the operator string where $A$ acts $l-1$ time intervals later than $B$. This auxiliary correlation function is related to the correlation function $\langle A(\tau) B \rangle$ through a convolution,
\beq
C_{AB}(l \: \delta\tau)=\int_{-\delta\tau}^{\delta\tau} d\tau\:\frac{\delta\tau-|\tau|}{\delta\tau^2}\Theta(\delta\tau-|\tau|)\langle A((l-1)\delta\tau+\tau) B\rangle.
\eeq
Instead of performing an approximate de-convolution of this relation~\cite{Michel07}, here we make use of the convoluted kernel
\beq
K_{\delta\tau}(\omega,\tau)=\int_{-\delta\tau}^{\delta\tau} d\tau' \:\frac{\delta\tau-|\tau'|}{\delta\tau^2}[e^{-(\tau+\tau')\omega}+e^{-(\beta-\tau-\tau')\omega}],
\eeq
which yields
\beq
K_{\delta\tau}(\omega,\tau)=\frac{(e^{\delta\tau\:\omega}-1)^2(1+e^{(2\tau-\beta)\omega})}{e^{(\delta\tau+\tau)\omega}\:\delta\tau^2\: \omega^2},
\eeq
in order
to directly relate the quantum Monte Carlo estimator $C_{AB}(l \: \delta\tau)$ to $S_{AB}(\omega)=\int dt \: e^{i\omega t} \langle A(t)B\rangle$ corresponding to the  real-time correlation function $\langle A(t)B\rangle$,
\beq
C_{AB}(l \: \delta\tau)=\int_0^\infty  \frac{d\omega}{2\pi} K_{\delta\tau}(\omega,\tau) S_{AB}(\omega).
\eeq
This relation holds true
without any approximation, irrespectively of the discretisation used in setting up the Monte Carlo estimator.
The final inversion problem is then solved using e.g. the stochastic analytic continuation approach, with the kernel $K_{\delta\tau}(\omega,\tau)$ for the present case.
As an example, we show in Fix.~\ref{fig:SBtau} the quantum Monte Carlo data for $S_B(\tau,\vec{k})=\langle B_\vec{k}(\tau) B_{-\vec{k}} \rangle$ at the $\Gamma$ point, $\vec{k}=0$, and a coupling ratio of $g=2.1$, along with the indicated asymptotic $1/\tau^4$-decay.

\section{Spin-wave theory}\label{App:SWT}

In this appendix, we provide details on both the  linear spin-wave theory calculations that we performed on the square-lattice bilayer as well as
on the higher-order spin-wave theory approach that we used.
In the classical limit of large spin $S$, the ground state of the Heisenberg model on the bipartite square lattice bilayer consists of an
antiferromagnetic arrangement of the spins within and between the layers, irrespectively of the coupling ratio $g$.
The classical spin structure can be parametrized by means of a single helix with pitch vector ${\bf Q}=(\pi,\pi,\pi)^\intercal$, such
that for a spin on site $l$,
\begin{equation}
 \vec{S}_l=S(0,0,e^{i{\bf Q \cdot r}_l}) \,.
\end{equation}
In this appendix, the index $l$ runs over all sites of the lattice, where $l=(i,\mu)$ is given by a rung  index $i$
plus a layer index $\mu$, as in the main text.
The quantum spin Hamiltonian is now rewritten expressing the
spin operators in the local basis of the classical spin orientations, $(x',y',z')$,
\begin{equation}
\begin{array}{l}
  S_{l}^x=e^{i{\vec{Q} \cdot \vec{r}}_l} S_{l}^{x'}\,, \\[3mm]
  S_{l}^y= S_{l}^{y'} \,, \\[3mm]
  S_{l}^z=e^{i{\vec{Q} \cdot \vec{r}}_l} S_{l}^{z'}\,.
\end{array}
\end{equation}
In this rotated frame, the classical ground state is thus ferromagnetic by construction.

\subsection{Linear spin-wave theory}

In the quantum model, the deviations from the classical order are expressed via the Holstein-Primakoff
representation of spin operators~\cite{Yosida96}. To first order in $1/S$, the expressions take the form
 \begin{equation}\label{eq:HP transformation TLAFM}
  \begin{array}{lll}
   S_{l}^{x'}&=& \displaystyle \frac{\sqrt{2S}}{2}(a_l^\dagga+a_{l}^\dagger)\,,\\[3mm]
   S_{l}^{y'}&=& \displaystyle \frac{\sqrt{2S}}{2i}(a_{l}^\dagga-a_{l}^\dagger)\,, \\[3mm]
   S_{l}^{z'}&=& S-n_{l}^\dagga\,,
   \end{array}
 \end{equation}
with $a_l^\dagga$ ($a^{\dagger}_l$) local bosonic annihilation (creation) operators, and $n_l=a^\dagger_l a_l^\dagga$.
In this bosonic representation the expression of the quantum Hamiltonian, truncated at the harmonic order, takes the following compact form in Fourier space
\begin{eqnarray}\label{eq:Hsw}
 H&=&-S(S+1)\left(2J+\frac{J'}{2}\right)N_S \nonumber \\
              && + S\sum_{\bf k}\left(a_{\bf k}^\dagger,a_{-{\bf k}}\right)
\left(\begin{array}{ll}
 A_{\bf k} &  B_{\bf k}\\
 B_{\bf k} &  A_{\bf k}
\end{array}\right)
\left(\begin{array}{l}
 a_{\bf k}\\
 a_{-{\bf k}}^\dagger
\end{array}\right),
\end{eqnarray}
where $N_S$ is the total number of lattice sites, and  the coefficients $A_{\bf k}$ and $B_{\bf k}$ are given by
\begin{equation}\label{eq:HPsw}
  \begin{array}{l}
 \displaystyle A_{\bf k}=2J+\frac{J'}{2}, \\[3mm]
 \displaystyle B_{\bf k}=-J(\cos k_x+\cos k_y)-\frac{J'}{2}\cos k_z,
  \end{array}
\end{equation}
with $k_z=0$ ($k_z=\pi$) for the symmetric (antisymmetric) sector.
Terms of order $S^2$ in Eq.~(\ref{eq:Hsw}) correspond to the classical energy and terms of order $S$  to the $1/S$ correction.
The quadratic Hamiltonian (\ref{eq:Hsw}) is diagonalized by introducing new bosonic quasiparticle operators via
\begin{equation}
 \begin{array}{l}
 a_{\bf k}^\dagga=u_{\bf k}^\dagga\alpha_{\bf k}^\dagga+v_{\bf k}^\dagga\alpha_{-{\bf k}}^\dagger, \\[3mm]
 a_{\bf k}^\dagger=u_{\bf k}^\dagga\alpha_{\bf k}^\dagger+v_{\bf k}^\dagga\alpha_{-{\bf k}}^\dagga,
\end{array}
\end{equation}
with $u_{\bf k}$ and $v_{\bf k}$ given by
\begin{eqnarray}
    u_{\bf k}&=&-\textrm{Sgn}(B_{\bf k})\left[\frac{1}{2}\left(\frac{A_{\bf k}}{\sqrt{A_{\bf k}^2-B_{\bf k}^2}}+1\right)\right]^\frac{1}{2},\nonumber\\
    v_{\bf k}&=&\left[\frac{1}{2}\left(\frac{A_{\bf k}}{\sqrt{A_{\bf k}^2-B_{\bf k}^2}}-1\right)\right]^{\frac{1}{2}}.
 \end{eqnarray}
The diagonal representation of  $H$ reads
\begin{equation}\label{eq:Hswdiag}
{H}=-S(S+1)\left(2J_\parallel+\frac{J_\perp}{2}\right)N_S +
              \sum_{\bf k}\omega_{\bf k}\left(\alpha_{\bf k}^\dagger \alpha^\dagga_{{\bf k}}+\frac{1}{2}\right),
\end{equation}
with the free magnon dispersion given by
\begin{equation}
\omega_{\bf k} =2S\sqrt{A_{\bf k}^2-B_{\bf k}^2}.
\end{equation}
The ground state of (\ref{eq:Hswdiag}) corresponds to the vacuum $|0\rangle$ of the Bogolyubov quasiparticles $\alpha$ and, for $S=1/2$,
we obtain a symmetric and an antisymmetric mode with dispersions
\begin{eqnarray}
\omega_{(k_x,k_y, 0)}  & = & 2 J \sqrt{\left[1 - \gamma(k_x,k_y)\right]\left[1+\gamma(k_x,k_y)+g/2\right]},\nonumber  \\
\omega_{(k_x,k_y,\pi)} & = & 2 J \sqrt{\left[1+ \gamma(k_x,k_y)\right]\left[1-\gamma(k_x,k_y)+g/2\right]},\nonumber
\end{eqnarray}
where
\begin{equation}\label{eq:appgamma}
\gamma(k_x,k_y)=\frac{1}{2}(\cos k_x+\cos k_y),
\end{equation}
 and $g=J'/J$.
They satisfy the relation $\omega_{\bf k}=\omega_{{\bf k}+{\bf Q}}$, which implies that the dispersion of the symmetric and antisymmetric
modes are identical up to a shift of the in-plane momentum by $(\pi,\pi)$.
The symmetric and antisymmetric modes soften respectively at $\vec{k}=0$ and $\vec{k}=\vec{Q}$.

We next calculate the single-mode contribution to the dynamical spin structure factor in terms of the  Bogolyubov quasiparticles.
To this end, we consider the operator
\begin{eqnarray}
S_{\vec{k}}^+ & = &\frac{1}{\sqrt{N_S}}\sum_{l}e^{i \vec{k} \cdot \vec{r}_l} S_{l}^+ \nonumber  \\
              & = & \frac{1}{\sqrt{N_S}}\sum_{l}e^{i \vec{k} \cdot \vec{r}_l} \left(e^{i \vec{Q}\cdot\vec{r}_l} S_{l}^{x'}+i S_{l}^{y'}\right) \nonumber  \\
              & = & \sqrt{\frac{S}{2}}\left(a_{\bf k}^\dagga+a_{{\bf k}+{\bf Q}}^\dagga+a_{-{\bf k}-{\bf Q}}^\dagger-a_{-{\bf k}}^\dagger\right).
\end{eqnarray}
In terms of the Bogolyubov quasiparticles, the effect of $S_{\bf k}^+$ onto the vacuum $|0\rangle$ is given by
\begin{equation}\label{eq:Sqonvacuum}
  S_{\bf k}^+|0\rangle =\sqrt{\frac{S}{2}}\left[ (v_{\bf k}-u_{\bf k})\alpha_{-{\bf k}}^\dagger+(v_{{\bf k}+{\bf Q}}+u_{{\bf k}+{\bf Q}})\alpha_{-{\bf k}-{\bf Q}}^\dagger\right]\!|0\rangle ,
\end{equation}
where we have used the property $u_{\bf k}=u_{-{\bf k}}$ and $v_{\bf k}=v_{-{\bf k}}$.
Inserting Eq.~(\ref{eq:Sqonvacuum}) into the Lehmann representation of the dynamical  spin structure factor,
and summing over the single-particle eigenstates of Eq.~(\ref{eq:Hswdiag}) gives
\begin{equation}\label{eq:DSFsw}
  S_S(\omega,\vec{k})=\mathcal{Z}_S(\vec{k})  \delta(\omega-\omega_{\bf k}),
\end{equation}
with the single-mode amplitude
\begin{equation}
\mathcal{Z}_S(\vec{k})=\frac{S}{2}\left(|v_{\bf k}-u_{\bf k}|^2+|v_{{\bf k}+{\bf Q}}+u_{{\bf k}+{\bf Q}}|^2 \right),
\end{equation}
where we have used the identity $\omega_{\bf k}=\omega_{{\bf k}+{\bf Q}}$.
Integrating $S_S(\omega,\vec{k})$ over $\omega$ yields the integrated spectral weight, $S_S(\vec{k})$, which within linear
spin-wave theory in the single-particle sector is thus given by the single-mode amplitude $\mathcal{Z}_S(\vec{k})$.
The integrated spectral weights  for  the symmetric and antisymmetric channels are thus given by
\begin{eqnarray}
\mathcal{Z}_S(k_x,k_y,0)&=&\frac{4S^2J(1-\gamma(k_x,k_y))}{\omega_{(k_x,k_y,0)}}\,, \\
\mathcal{Z}_S(k_x,k_y,\pi)&=&\frac{4S^2J(1-\gamma(k_x,k_y)+g/2)}{\omega_{(k_x,k_y,\pi)}}\,.
\end{eqnarray}
Thus we recover the limiting behaviour obtained numerically: the amplitude of the symmetric mode at $\vec{k}=0$ vanishes linearly,
while the amplitude of the antisymmetric mode at the antiferromagnetic  Bragg peak position $\vec{k}=\vec{Q}$ diverges like
$\sim1/ | \vec{k}-\vec{Q} |$, namely,
\begin{eqnarray}
 \mathcal{Z}_S(k,k,0)   & \approx & \frac{J S k} {\sqrt{J(4J + J')}}\,,\quad k\rightarrow 0\,, \\
 \mathcal{Z}_S(k,k,\pi) & \approx & \frac{S \sqrt{J (4J + J') } } {J(k-\pi)}\,,\quad k\rightarrow \pi\,,
\end{eqnarray}
respectively.

\subsection{Higher-order spin-wave theory}

As discussed in the main text, the agreement with the quantum Monte Carlo data is improved if the spin-wave expansion is extended
beyond the harmonic order. The first step is to truncate the Holstein-Primakoff expansion to next-to-leading order. This amounts to keeping, in the expression of the spin operators in the rotated frame $(x^\prime,y^\prime,z^\prime)$, terms
of order $S^{-1/2}$, such that
 \begin{equation}\label{eq:HP transformation TLAFM 2}
  \begin{array}{lll}
   S_l^{x'}&=& \displaystyle \frac{\sqrt{2S}}{2}(a_l^\dagga+a_l^\dagger)-\frac{1}{4\sqrt{2S}}(n_l^\dagga a_l^\dagga+a_l^\dagger n_l^\dagga)\,,\\[3mm]
   S_l^{y'}&=& \displaystyle \frac{\sqrt{2S}}{2i}(a_l^\dagga-a_l^\dagger)-\frac{1}{4i\sqrt{2S}}(n_l^\dagga a_l^\dagga-a_l^\dagger n_l^\dagga)\,,\\[3mm]
   S_l^{z'}&=& S-n_l^\dagga \,.
   \end{array}
 \end{equation}

Given the collinear nature of the classical state about which we perform the expansion,
the terms beyond the usual harmonic contribution will be of quartic order in
bosonic operators. To this level of approximation, we are left with a Hamiltonian
\begin{equation}
{H}={H}^{(0)}+{H}^{(2)}+{H}^{(4)}+\mathcal{O}(1/S)\,,
\end{equation}
where ${H}^{(0)}$, ${H}^{(2)}$ and ${H}^{(4)}$ are respectively the classical energy contribution $\sim S^2$, the harmonic fluctuation Hamiltonian $\sim S$,
and the collection of all quartic interaction terms $\sim 1$. Lastly, $\mathcal{O}(1/S)$ denotes all the remaining terms in the expansion which are of order $1/S^\alpha$ with $\alpha\geq1$ and which we neglect.
The expression for the sum ${H}^{(0)}+{H}^{(2)}$ is given in Eq.~(\ref{eq:Hsw}), while ${H}^{(4)}$ is obtained as
\begin{eqnarray}
&{H}^{(4)}&\nonumber\\
&=&\frac{J}{2}\sum_{l,{\bf d}\in\{\pm {\bf x},\pm {\bf y}\}}{\frac{1}{4}\left[(n_{l}^\dagga+n_{l+{\bf d}}^\dagga)a_{l}^\dagga a_{l+{\bf d}}^\dagga +\textrm{h.c.}\right]-n_{l}^\dagga n_{l+{\bf d}}^\dagga}\nonumber \\[3mm]
                  & & +\frac{J'}{4}\sum_{l,{\bf d}\in\{\pm{\bf z}\}}{\frac{1}{4}\left[(n_{l}^\dagga+n_{l+{\bf d}}^\dagga)a_{l}^\dagga a_{l+{\bf d}}^\dagga +\textrm{h.c.}\right]-n_{l}^\dagga n_{l+{\bf d}}^\dagga}\,.\nonumber
\end{eqnarray}

To deal with this hamiltonian, the simplest approximation consists in performing a mean-field decoupling of the quartic boson terms, and to evaluate the expectation values in the harmonic ground state. This yields an effective harmonic contribution ${H}_{\textrm{eff}}^{(4)}\sim 1$ which adds up to the quadratic terms ${H}^{(2)}\sim S$. Given the main objective of this calculation, which is just to demonstrate that the qualitative tendency is correct, we have not attempted to go beyond this simple treatment and to make the calculation of the expectation values self-consistent. This more sophisticated and significantly heavier approach, known as self-consistent spin-wave theory, includes higher-order corrections in an approximate way and would probably further improve the agreement, but it would anyway not allow to describe the quantum phase  transition.

In Fourier space the sum $H_\textrm{eff}={H}^{(0)}+{H}^{(2)}+{H}_{\textrm{eff}}^{(4)}$ has the same structure as Eq.~(\ref{eq:Hsw}), with
\begin{eqnarray}\label{eq:Hsw4}
{H}_\textrm{eff}&=&-S(S+1)\left(2J+\frac{J'}{2}\right)N_S \nonumber \\
&+& S\sum_{\bf k}\left(a_{\bf k}^\dagger,a_{-{\bf k}}\right)
\left(\begin{array}{ll}
 A_{\bf k}^\textrm{eff} &  B_{\bf k}^\textrm{eff}\\
 B_{\bf k}^\textrm{eff} &  A_{\bf k}^\textrm{eff}
\end{array}\right)
\left(\begin{array}{l}
 a_{\bf k}\\
 a_{-{\bf k}}^\dagger
\end{array}\right)
+\textrm{C}\,, \nonumber
\end{eqnarray}
where C is an additional constant of order $\mathcal{O}(1)$ and $A_{\bf k}^\textrm{eff}, B_{\bf k}^\textrm{eff}$ are the coefficients of Eq.~(\ref{eq:HPsw}),
evaluated at the effective couplings $J_\textrm{eff}$ and $ J'_\textrm{eff}$
(i.e., $A_{\bf k}^\textrm{eff}=A_{\bf k}(J_\textrm{eff},J'_\textrm{eff})$ and $B_{\bf k}^\textrm{eff}=B_{\bf k}(J_\textrm{eff},J'_\textrm{eff})$), which are given by
\begin{equation}\label{eq:Jeff}
 J_\textrm{eff}=J\left(1+\frac{\Delta-n}{S}\right), \quad J'_\textrm{eff}=J'\left(1+\frac{\Delta'-n}{S}\right),
\end{equation}
where $n$, $\Delta$ and $\Delta'$ are the following two-body averages computed in the harmonic approximation
\begin{equation}
\begin{array}{c}
 n=\langle a_{l}^\dagger a_{l}^\dagga \rangle \,,\\[3mm]
 \Delta=\langle a_{l}^\dagga a_{l+{\bf d}}^\dagga \rangle \quad \textrm{for } {\bf d}\in\{\pm{\bf x},\pm{\bf y}\} \,,\\[3mm]
 \Delta'=\langle a_{l}^\dagga a_{l+{\bf d}}^\dagga \rangle \quad \textrm{for } {\bf d}\in\{\pm{\bf z}\}\,.
 \end{array}
\end{equation}
Note that all other two-body averages vanish,  $\langle a_{l}^\dagger a_{l+{\bf d}}^\dagga \rangle=0$, and $\langle a_l^\dagga a_l^\dagga \rangle=0$.
The next-to-leading-order term of the spin-wave expansion thus has the effect of renormalizing the coupling ratio $g=J'/J$ via
\begin{equation}
g   \longmapsto g_\textrm{eff}=g \frac{S+\Delta'-n}{S+\Delta-n}\,.
\end{equation}

\begin{figure}[t]
 \centering
 \includegraphics[width=\columnwidth]{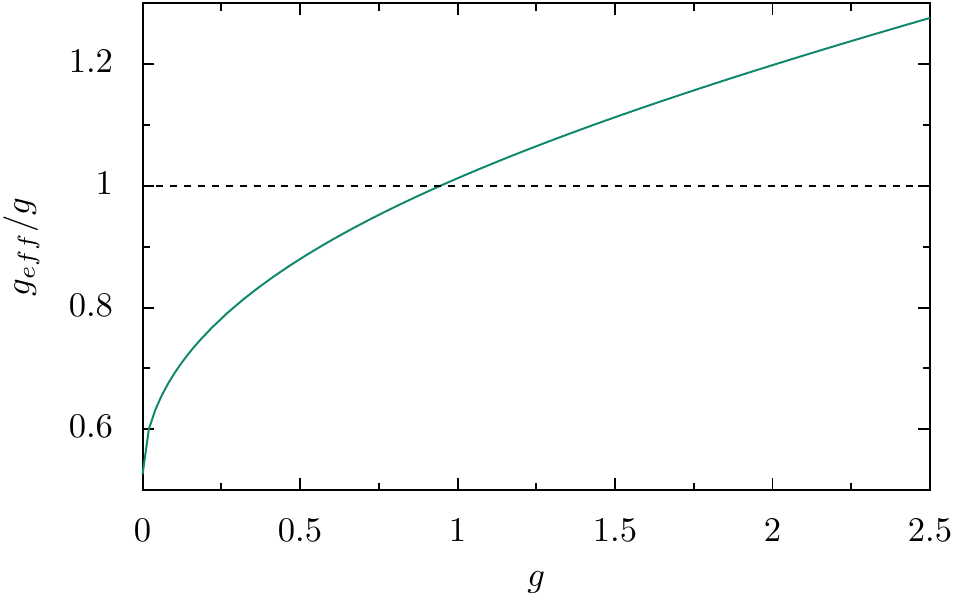}
\caption{(Color online) Effective coupling ratio $g_\textrm{eff}$ as a function of $g$ for $S=1/2$.}
\label{fig:Multiplicative_coeff}
\end{figure}

\begin{figure}[t]
 \centering
 \includegraphics[width=\columnwidth]{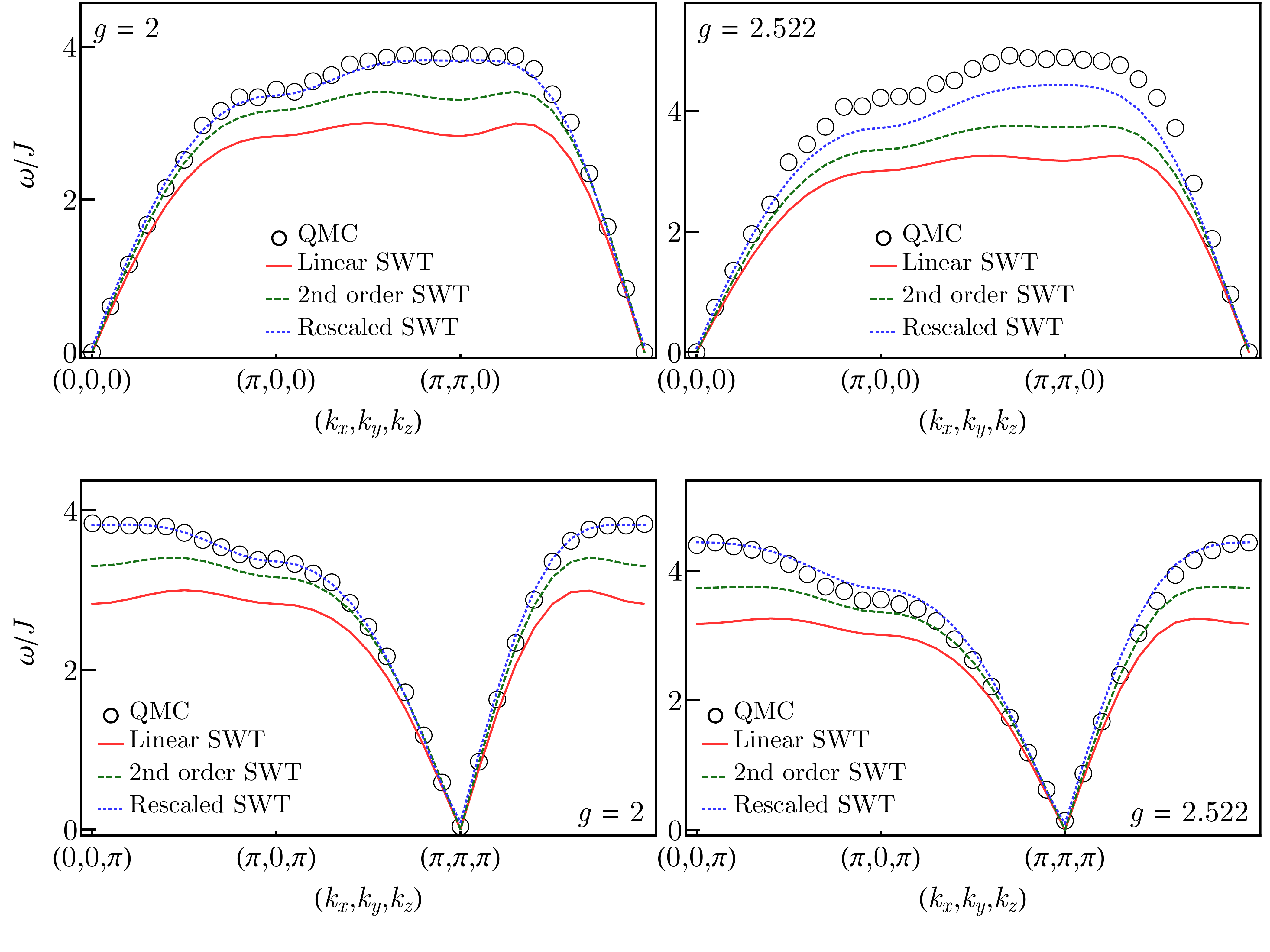}
\caption{(Color online) Plots of the dynamical spin structure factor at different ratios $g=J'/J$ inside the antiferromagnetically ordered region for the spin-half Heisenberg model on the square lattice bilayer.
The panels compare the quantum Monte Carlo (QMC) results to both the linear and 2nd order spin-wave theory results as well as to further rescaled
effective coupling ratios, which for $g=2$ and $g=2.522$ are obtained as $g_\textrm{eff}=1.8244g$ and $g_\textrm{eff}=1.95g$ respectively.}
\label{fig:Multiplicative_coeff_comp}
\end{figure}

Depending on the values of the bare couplings  $J$ and $J'$, the multiplicative coefficient ranges from values smaller than one to values bigger than one,
as shown in Fig.~\ref{fig:Multiplicative_coeff}. This figure suggests that for ratios $g \geq1$, spin-wave interactions yield an effective ratio $g_\textrm{eff}$
which is larger than the bare value.
We conjecture that this renormalization will be even bigger if higher-order spin-wave interactions are taken into account.
For example, we find that the quantum Monte Carlo results at $g=2$ are rather well fitted by the spin-wave dispersion if the effective ratio, $g_\textrm{eff}$, used is even larger than the one obtained from this improved spin-wave theory, as shown in
Fig.~\ref{fig:Multiplicative_coeff_comp}. This figure also shows that for $g=2.522$, an almost similarly good fit can be obtained.

\section{Perturbation theory in $1/g$}\label{App:PT}

For vanishing intralayer coupling $J=0$, i.e. in the limit $1/g=0$, the ground state of the bilayer spin system is given by the direct
product of singlets on the interlayer bonds (rungs).
We denote this state by $|S\rangle=\bigotimes_i |s\rangle_i$, where $|s\rangle_i$ is the singlet on rung $i$,
\begin{equation}
 |s\rangle_i=\frac{1}{\sqrt{2}}\left(|\!\uparrow\rangle_{i,1}|\!\downarrow\rangle_{i,2}-|\!\downarrow\rangle_{i,1}|\!\uparrow\rangle_{i,2}\right)\,.
\end{equation}
To treat the intralayer coupling $J$ as a weak perturbation in the small $1/g$ regime, we separate the total Hamiltonian $H$ of the bilayer model in
the interlayer and intralayer parts $H_\perp$ and $H_\parallel$, which scale respectively with $J'$ and $J$.
$|S\rangle$  is thus the ground state of $H_\perp$.
Treating $H_\parallel$ as a weak perturbation to $H_\perp$, the ground state to first order in $1/g$ is found to be
\begin{eqnarray}\label{eq:LargeJperpgs}
 |GS\rangle=|S\rangle-\frac{1}{4g}&&\sum_{i,\vec{d}\in\{\vec{x},\vec{y}\}}\big[ \: |t_0\rangle_i|t_0\rangle_{i+\vec{d}}\\
 & &-|t_1\rangle_{i}|t_{-1}\rangle_{i+\vec{d}}-|t_{-1}\rangle_{i}|t_{1}\rangle_{i+\vec{d}} \big]\,,\nonumber
 \end{eqnarray}
where the state $|t_n\rangle_i|t_m\rangle_{i+d}$ is the direct product of the triplets $|t_n\rangle$ and $|t_m\rangle$ on rungs $i$ and $i+\vec{d}$ times
the direct product of singlets on all other rungs (which are thus implicit in this notation). To study the dynamical spin structure factor, we consider the following  local interlayer bond operators
which relate to the antisymmetric  ($k_z=\pi$) and the symmetric ($k_z=0$) modes:
\begin{eqnarray}
 S_{i}^\textrm{AS}&=&\frac{1}{\sqrt{2}}\left(S_{i,1}^+ - S_{i,2}^+\right),\\
  S_{i}^\textrm{S}&=&\frac{1}{\sqrt{2}}\left(S_{i,1}^+ + S_{i,2}^+\right).
\end{eqnarray}
These operators act on the eigenstates of a rung as follows:
\begin{equation}\label{eq:definition of modes}
\begin{array}{lll}
  \begin{array}{l}
       S_{i}^\textrm{AS}|s\rangle_i=-|t_1\rangle_i \\[3mm]
       S_{i}^\textrm{AS}|t_1\rangle_i=0 \\[3mm]
       S_{i}^\textrm{AS}|t_0\rangle_i=0 \\[3mm]
       S_{i}^\textrm{AS}|t_{-1}\rangle_i=|s\rangle_i
      \end{array}
& \quad &
  \begin{array}{l}
       S_{i}^\textrm{S}|s\rangle_i=0 \\[3mm]
       S_{i}^\textrm{S}|t_1\rangle_i=0 \\[3mm]
       S_{i}^\textrm{S}|t_0\rangle_i=|t_1\rangle_i  \\[3mm]
       S_{i}^\textrm{S}|t_{-1}\rangle_i=|t_0\rangle_i
      \end{array}
\end{array}
\end{equation}
We first consider the antisymmetric sector.
We start by computing the effect of 
\begin{equation}\label{eq:FTASmode}
S_{\bf k}^{AS}=\frac{1}{\sqrt{N}}\sum_i S_{i}^{AS} e^{i {\bf k}\cdot {\bf r}_i}
\end{equation}
on the approximate ground state Eq.~(\ref{eq:LargeJperpgs}) (where $\vec{k}$ is a two-dimensional vector and $N$ is the number of interlayer bonds).
According to (\ref{eq:definition of modes}) this operator promotes a singlet to a triplet, and
thus the antisymmetric mode captures the dynamics of triplet excitations.
Restricting ourselves to the single triplon sector, we obtain:
\begin{equation}\label{eq:AS operator on GS}
 S_{\bf k}^{AS}|GS\rangle=\left(-1+\frac{1}{g}\gamma(k_x,k_y)\right)|t_1\rangle_{\bf k}\,,
\end{equation}
with
\begin{equation}
|t_1\rangle_{\bf k}=\frac{1}{\sqrt{N}}\sum_i e^{i {\bf k}\cdot{\bf r}_i} |t_1\rangle_i\,,
\end{equation}
and $\gamma(k_x,k_y)$ defined in Eq.~(\ref{eq:appgamma}).
To compute the matrix elements entering the expression of the dynamical spin structure factor in the Lehmann representation,
the knowledge of the full spectrum is required. Here, we will consider the perturbative eigenstates.
Since we consider only the single triplet component of $S_{\bf k}^{AS}|GS\rangle$ (see Eq.~\ref{eq:AS operator on GS}),
we can restrict ourselves to the eigenstates of the effective Hamiltonian
\begin{equation}\label{eq:Heff}
 H^\textrm{eff}=H_\perp+P_1H_\parallel P_1+P_1H_\parallel S H_\parallel P_1\,
\end{equation}
where $P_1$ is the projector onto the single triplet degenerate manifold, and $S=(1-P_1)/(E_{1t}-H_\perp)$,
$E_{1t}$ denoting the energy of the single triplet degenerate manifold.
The states $|t_1\rangle_{\bf k}$ are found to be  eigenstates of ${H}^\textrm{eff}$
with eigenvalues
\begin{equation}
\begin{array}{ll}
 H^\textrm{eff}|t_1\rangle_{\bf q}=&\big\{(-\frac{3}{4}J' N+J')+2J\gamma(k_x,k_y)\\[3mm]
                                   &-\frac{J^2}{J'}(\frac{3}{4}(N-2)+\frac{1}{2}(4\gamma(k_x,k_y)^2-1)\big\} |t_1\rangle_{\bf k}\, .
\end{array}
\end{equation}
A similar calculation allows us to compute the ground state energy to the same order in $J/J'$,
\begin{equation}
E_0= -\frac{3}{4}J' N-\frac{3}{4}\frac{J^2}{J'} N\,.
\end{equation}
Thus, overall we obtain for $\vec{k}=(k_x,k_y,\pi)^\intercal$, i.e., in the antisymmetric channel, the single-triplon contribution to the dynamical spin structure factor as
\begin{equation}\label{eq:Dynamical structure factor antisymmetric mode 1}
 S_S(\omega, \vec{k})=\mathcal{Z}_S(\vec{k}) \delta(\omega-\omega_{\bf k})\, , 
 \end{equation}
with the spectral weight amplitude to first order in $1/g$  given by
\begin{equation}\label{eq:Amplitude dynamical structure factor antisymmetric mode}
\mathcal{Z}_S(\vec{k}) =1-\frac{2}{g}\gamma(k_x,k_y)\,.
\end{equation}
The single triplon dispersion, correct up to quadratic order, is given by
\begin{equation}\label{eq:Dynamical structure factor antisymmetric mode 2}
\omega_{\bf k}=J'+2J\gamma(k_x,k_y)-2\frac{J^2}{J'}\left[\gamma(k_x,k_y)^2-1\right]\, .
\end{equation}
Equation (\ref{eq:Amplitude dynamical structure factor antisymmetric mode}) shows that the amplitude of the dynamical spin structure factor
for the antisymmetric mode is of order 1.
The momentum modulations of the amplitude are an effect of order $1/g$, with the largest amplitude for $\vec{k}=(\pi,\pi,\pi)^\intercal$
and the lowest at $\vec{k}=(0,0,\pi)^\intercal$.
These results compare well with the quantum Monte Carlo structure factor, as discussed in the main text (see Fig.~\ref{fig:SScomp2}).

\begin{figure}[t]
\centering
\includegraphics[width=\columnwidth]{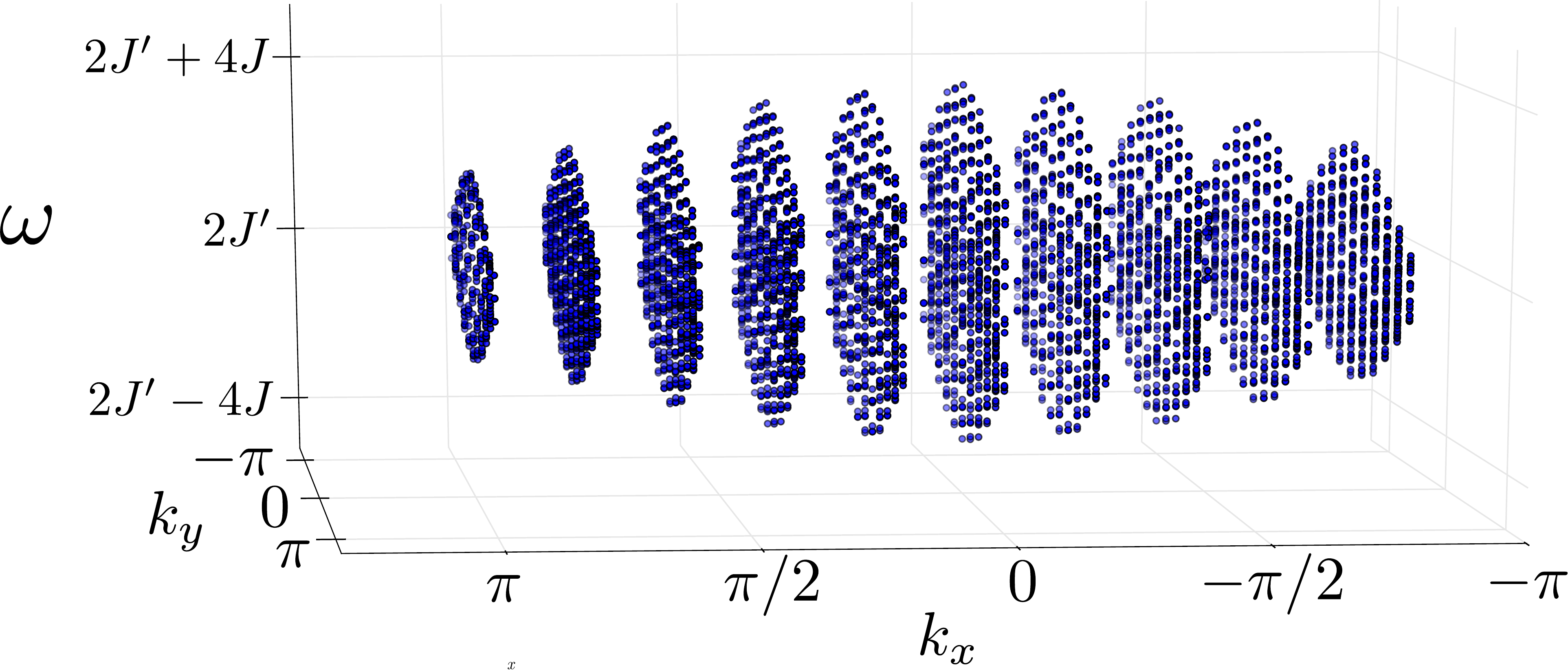}
\caption{Spectrum of $H^\textrm{eff}_2$ within the two-triplet sector for a system size $L=10$.
Energies are measured with respect to the ground state energy.}
\label{fig:Full Continuum spectrum}
\end{figure}

Next, we  turn to the calculation of the symmetric channel.
We start by computing, for $\vec{k}=(k_x,k_y)^\intercal$, the effect of
\begin{equation}\label{eq:FT S mode}
S_{\bf k}^{S}=\frac{1}{\sqrt{N}}\sum_i S_{i}^{S} e^{i {\bf k}\cdot{\bf r}_i}
\end{equation}
on the approximate ground state in Eq.~(\ref{eq:LargeJperpgs}).
According to Eqs.~(\ref{eq:definition of modes}), this operator cancels the singlet component of the wave function. The finite contributions come from the triplet component, and are of order $1/g$,
\begin{equation}\label{eq:Sym mode on gs}
 S_{\bf k}^{S}|GS\rangle=-\frac{1}{4g}\sum_{\vec{d}\in\{\vec{x},\vec{y}\}}\left(|t_1t_0\rangle_{{\bf k},{\bf d}}-|t_0t_1\rangle_{{\bf k},{\bf d}}\right)\left(1-e^{i{\bf k}\cdot \vec{d}}\right)\,,
\end{equation}
where we have introduced
\begin{eqnarray}
|t_1t_0\rangle_{{\bf k},{\bf d}}&=&\frac{1}{\sqrt{N}}\sum_{i}{e^{i{\bf k}\cdot\vec{r}_i}}|t_1\rangle_{i}|t_0\rangle_{i+\vec{d}}\,,\\
|t_0t_1\rangle_{{\bf k},{\bf d}}&=&\frac{1}{\sqrt{N}}\sum_{i}{e^{i{\bf k}\cdot\vec{r}_i}}|t_0\rangle_{i}|t_1\rangle_{i+\vec{d}}\,,
\end{eqnarray}
the Fourier transforms of neighboring pairs of triplets with zero relative momenta.
Similarly as before, to compute the matrix elements  required for the dynamical spin structure factor,
we will consider the perturbative eigenstates. Since, to order $1/g$, the states $S_{\bf q}^{S}|GS\rangle$ contain two triplets, $|t_0\rangle$
and  $|t_1\rangle$, we can restrict the calculation to the eigenbasis of the effective Hamiltonian
\begin{equation}\label{eq:Heff 2}
 H_2^\textrm{eff}=H_\perp+P_2H_\parallel P_2,
\end{equation}
where $P_2$ is the projector onto the two triplet degenerate subspace.
The effective Hamiltonian describes the two-body dynamics of two triplets ($|t_0\rangle$ and $|t_1\rangle$) in a sea of singlets.
The full diagonalisation of $H_\parallel$ inside the two-triplet sector is non trivial, and we performed it numerically.
We considered a $L=10$ lattice, yielding a two-triplet Hilbert space of dimension 9900.
We performed a full diagonalisation, resolving the energies in momentum space.
The spectrum of $H_2^\textrm{eff}$ forms a continuum in the thermodynamic limit; for our lattice size we have access to 100 values of the momentum
in the whole Brillouin zone, and for each momentum, we get 99 eigenstates with different energies.
The spectrum, with energies relative to the energy of the ground state, disperses around $2J^\prime$, the energy cost of
promoting two singlets into triplets. It is shown in Fig.~(\ref{fig:Full Continuum spectrum}).
We do not obtain an explicit expression for the dynamical spin structure factor in the symmetric channel
since the eigenbasis
and eigenenergies
of ${H}_2^\textrm{eff}$ are obtained only numerically.
Instead, for each value of the momentum, we computed the overlap between $S_{\bf k}^{S}|GS\rangle$
and the eigenbasis of ${H}_2^\textrm{eff}$. Thus, for each momentum, there is a distribution of amplitudes at different energies.
This distribution of amplitudes is shown in Fig.~\ref{fig:SSsympt} of the main text.
Based on our analysis,
a few remarks can be made about the dynamical spin structure factor associated to the symmetric mode:
(i) It can be seen analytically that at zero momentum the amplitude of the dynamical spin structure factor in the symmetric channel vanishes
(this may also be seen in Fig.~\ref{fig:SSsympt}).
In fact, from Eq.~(\ref{eq:Sym mode on gs}) one obtains $S_{{\bf k}=0}^{S}|GS\rangle=0$.
(ii)  At momentum $\vec{k}=(\pi,\pi)^\intercal$, one can verify that $S_{{\bf k}=(\pi,\pi)^\intercal}^{S}|GS\rangle$ is an eigenstate of ${H}_2^\textrm{eff}$
with energy equal to $2J'-J/2$ when counted from the energy of the groundstate.
This explains why at $k=(\pi,\pi)^\intercal$ all the amplitude is concentrated at $\omega=2J'-J/2$ in Fig.~\ref{fig:SSsympt}.
(iii)
The amplitude of the symmetric mode is of order $(1/g)^2$, while that of the antisymmetric mode is of order $1$.
Thus, the signal of the symmetric mode vanishes in the limit $1/g\rightarrow 0$ (see Fig.~\ref{fig:SS}).


\section{Bond-bond correlations from the $1/d$ expansion}\label{App:BOT}

In this appendix, we provide details of the calculation of the dispersion and the single-particle weight in the bond-bond correlations, which are related to the Higgs-peak contribution of the dynamical singlet structure factor $S_B(\omega, \vec{k})$, using the bond-operator-based $1/d$ expansion \cite{Joshi15a, Joshi15b}.

The starting point is a model of coupled dimers described by Eq.~\eqref{hh}, but now on a hypercubic lattice in $d$ spatial dimensions, such that the square-lattice bilayer corresponds to $d=2$. To obtain a non-trivial large-$d$ limit, it is crucial to properly scale the ratio between the interdimer coupling $\jp$ and the intradimer coupling $\jpl$. To this end we introduce the ratio
\begin{equation}
\label{q}
\qq=\jpl d/\jp
\end{equation}
as our tuning parameter, with the notion that $\qq$ is kept fixed upon taking the $d\to\infty$ limit.
We refer the reader to Ref.~\onlinecite{Joshi15a} for an extended discussion of the $1/d$ expansion approach and the general formalism. In the following,  we will restrict ourselves to $T=0$ calculations. Further, most explicit expressions will be restricted to the leading order in $1/d$. Note that to this order (i.e. at the harmonic level) of approximation, the quantum critical point is located at $\qq=\qq_c = 1/2$.

As shown in detail in Refs.~\onlinecite{Joshi15a, Joshi15b}, the bond-operator approach yields a triply degenerate triplon (spin-$1$ excitation) mode in the quantum disordered phase (i.e. for $\qq < \qq_c$) while in the antiferromagnetically ordered phase (i.e. for $\qq > \qq_c$) one obtains two degenerate transverse modes (the Goldstone modes) and a longitudinal amplitude mode.
In view of its further usage we quote here the leading-order result~\cite{Joshi15b}
for the dispersion of the longitudinal mode (valid for $\qq \geq 1/2$):
\begin{equation}
\wkz = 2\jp\qq \sqrt{1 + \frac{\gk}{4\qq^2}} \,,
\end{equation}
where $\vec{k} = (k_1, k_2, ..., k_d)^\intercal$ and
\begin{equation}
\gk = \frac{1}{d} \sum_{i=1}^{d} \cos k_i \,.
\end{equation}

\begin{figure*}[t]
\centering
\includegraphics[width=\textwidth]{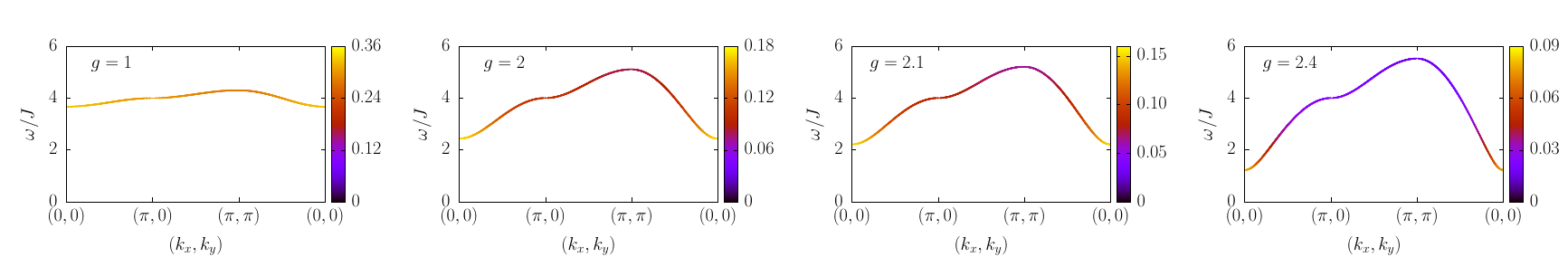}
\caption{Spectral weight distribution in the dynamic singlet structure factor for different values of $g$, as obtained from the bond-operator theory using the mapping in Eq.~(\ref{eq:relativemapping}).}
\label{fig:BOTcombo}
\end{figure*}

For the interlayer bond-bond correlations we consider the susceptibility
\begin{equation}
\label{bbc_def}
\chi_{\bo}(\omega,\vec{k}) = -i \int_{-\infty}^{\infty} dt e^{i \omega t} \langle T_t \bo_{\vec{k}}(t) \bo_{-\vec{k}}(0) \rangle \,,
\end{equation}
where $T_t$ is the time-ordering operator, and
\begin{align}
\bo_i &= \vec{S}_{i1} \cdot \vec{S}_{i2} \,, \\
\bo_{\vec{k}} &= \frac{1}{\sqrt{N}} \sum_{i} \bo_i e^{-i \vec{k} \cdot \vec{r}_i} \,.
\end{align}
By the fluctuation-dissipation theorem, the imaginary part of the susceptibility $\chi_B$ is proportional to the structure factor $S_B$ as considered in the main text.
Expressed in terms of bond operators (we follow the notation of Ref.~\onlinecite{Joshi15a}), we obtain
\begin{equation}
\bo_i = \sum_{\alpha=x,y,z} t_{i\alpha}^{\dagger} t_{i\alpha} - \frac{3}{4} \,.
\end{equation}
In the disordered phase,  the $t_{i\alpha}$ ($t_{i\alpha}^{\dagger}$) are annihilation (creation) operators for local spin-$1$ excitations, introduced within the  bond-operator formulation.
In the following, we work in the single-mode approximation (see Ref.~\onlinecite{Joshi15b} for  details), i.e., we will ignore multi-mode contributions to the bond-bond correlation function which would cause continua in the spectrum. As a result, we neglect terms containing more than two triplon operators in the operator product $\bo_{\vec{k}} \bo_{-\vec{k}}$. We are then lead to the following expression for the interlayer bond susceptibility in the disordered phase:
\begin{equation}
\label{interpl_dis}
\chi_{B}^{dis} (\omega,\vec{k}) = N \left[ \frac{9}{16} - \frac{9}{2} R_2 + \mathcal{O}(1/d^{2}) \right] \delta_{\vec{k},0} \delta(\omega) \,,
\end{equation}
where $R_2 = \langle t_{i\alpha}^{\dagger} t_{i\alpha} \rangle$ up to $\mathcal{O}(1/d)$. We thus do not obtain any spectral weight in this sector apart from a Bragg-like peak at $\omega = 0$. (We note that a two-triplon continuum appears at order $1/d$.)

In the ordered phase, we performed the  calculations to leading order in $1/d$, i.e., at the harmonic level. Here, we use generalized triplon operators $\tilde{t}$ and $\tilde{t}^{\dagger}$, obtained from the
$t$ and $t^{\dagger}$ operators after a suitable rotation in  Hilbert space~\cite{Joshi15b}. Note that the $\tilde{t}_x$ and $\tilde{t}_y$ operators refer to the two degenerate transverse modes, while the $\tilde{t}_z$ operator corresponds to the longitudinal mode. Furthermore, we define the condensate parameter
\begin{equation}
\lambda = \sqrt{\frac{2\qq-1}{2\qq+1}} \,,
\end{equation}
corresponding to the condensation of one of the triplon modes in the ordered phase. We then obtain
\begin{align}
\bo_i &= \vec{S}_{1i} \cdot \vec{S}_{2i} \nonumber \\
&= -\frac{3}{4} + \tilde{t}_{ix}^{\dagger} \tilde{t}_{ix} + \tilde{t}_{iy}^{\dagger} \tilde{t}_{iy} \nonumber \\
&+ \frac{1}{1+\lambda^{2}} \left[ \tilde{t}_{iz}^{\dagger} \tilde{t}_{iz}  + \lambda^{2} P_i + \lambda e^{i \vec{Q}\cdot \vec{r}_i} (\tilde{t}_{iz}^{\dagger} + \tilde{t}_{iz}) \right] \nonumber  \\
&= -\frac{3}{4} + \frac{\lambda^{2}}{1+\lambda^{2}} + \frac{\lambda}{1+\lambda^{2}} e^{i \vec{Q}\cdot \vec{r}_i} (\tilde{t}_{iz}^{\dagger} + \tilde{t}_{iz})  + \tilde{t}_{iz}^{\dagger} \tilde{t}_{iz}   \nonumber  \\
&+ \frac{\tilde{t}_{ix}^{\dagger} \tilde{t}_{ix} + \tilde{t}_{iy}^{\dagger} \tilde{t}_{iy}}{1 + \lambda^{2}} \,,
\end{align}
where $\vec{Q}=(\pi, \pi, ... )^\intercal$ and in the last equation we have used the singlet projector
$P_i = 1-\sum_{\alpha} \tilde{t}_{i\alpha}^{\dagger} \tilde{t}_{i\alpha}$ (with $\alpha=x,y,z$). This gives
\begin{align}
\bo_{\vec{k}} &= \sqrt{N} \left[ -\frac{3}{4} + \frac{\lambda^{2}}{1+\lambda^{2}} \right]\delta_{\vec{k},0}
+ \frac{\lambda}{1+\lambda^{2}} (\tilde{t}_{\vec{k}-\vec{Q},z}^{\dagger} + \tilde{t}_{-\vec{k}+\vec{Q},z}) \\ \nonumber
&+ \frac{1}{\sqrt{N}} \sum_{\vec{q}} \left[ \tilde{t}_{\vec{q},z}^{\dagger} \tilde{t}_{\vec{q}-\vec{k},z}
+ \frac{\tilde{t}_{\vec{q},x}^{\dagger} \tilde{t}_{\vec{q}-\vec{k},x} + \tilde{t}_{\vec{q},y}^{\dagger} \tilde{t}_{\vec{q}-\vec{k},y}}{1+\lambda^{2}}\right] \,.
\end{align}
As noted above,  we now ignore the last term in the above equation since it will either contribute to the continuum or give a $1/d$ correction to the single-particle peak. Hence, we do not obtain a single-mode contribution from the transverse modes. We are thus left with
\begin{align}
\label{interpl_ord}
\chi_{B}^{ord} &(\omega,\vec{k}+\vec{Q}) = N \left[ -\frac{3}{4} + \frac{\lambda^{2}}{1+\lambda^{2}} \right]^{2} \delta_{\vec{k},0} \delta(\omega) \\ \nonumber
&+ \frac{\lambda^{2}}{1+\lambda^{2}} \left[ \mathcal{G}_{N,z} (\omega, \vec{k}) + \mathcal{G}_{N,z} (-\omega,\vec{k}) \right. \\ \nonumber
&\left. + \mathcal{G}_{A,z} (\omega,\vec{k}) + \mathcal{G}_{A,z} (-\omega, \vec{k})\right] \\
&= N \left[ -\frac{3}{4} + \frac{\lambda^{2}}{1+\lambda^{2}} \right]^{2} \delta_{\vec{k},0} \delta(\omega) \\ \nonumber
&+ \frac{\lambda^{2}}{1+\lambda^{2}} (u_{\vec{k},z} + v_{\vec{k},z})^{2} \left[ \frac{1}{\omega - \wkz} - \frac{1}{\omega + \wkz} \right] \,,
\end{align}
where we have used the fact that $\vec{k}$ resides within the first Brillouin zone, and  the fact that $\vec{k} \pm 2\vec{Q}$ is equivalent to $\vec{k}$.
In the above expression, $\mathcal{G}_{N}$ and $\mathcal{G}_{A}$ denote the zero-temperature normal and anomalous Green's function, respectively.
Also, $u_{\vec{k},z}$ and $v_{\vec{k},z}$ are Bogoliubov coefficients, used
in diagonalising the Hamiltonian to leading order in $1/d$, and at this point we refer to Ref.~\onlinecite{Joshi15b} for their explicit expressions.
We can now read-off the mode weight corresponding to $\wkz$ from the above expression,
\begin{align}
\mathcal{Z}_B({\vec{k}+\vec{Q}}) &= \frac{\lambda^{2}}{1+\lambda^{2}} (u_{\vec{k},z} + v_{\vec{k},z})^{2} 
= \frac{2\qq - 1}{2 \sqrt{4 \qq^{2} + \gk}} \,, 
\end{align}
As discussed in Sec. III, we need to account for the shift in the location of the quantum critical point  within the bond-operator approach from the actual value, when comparing
the above results to the quantum Monte Carlo data.
In Fig.~\ref{fig:BOTcombo}, we show the single-mode contribution to the spectral weight atop the corresponding dispersion relation for different values of $g$, using the mapping in Eq.~(\ref{eq:relativemapping}) to the corresponding values of $q$ used in the above formula. We recall that, in addition to this single-mode contribution, a two-magnon continuum appears at leading order in $1/d$ which we have not determined here.
Finally, we note that within the current approach, one can similarly calculate the quasi-particle contribution to the bond-bond correlations  among the  intralayer bonds. Similar to the case of the spin-spin correlations, one can in this case distinguish between a symmetric and antisymmetric channel with respect to layer inversion symmetry. For the symmetric channel, we obtain again a single-mode contribution like in the interlayer case considered explicitly above, however of reduced  spectral weight. In the antisymmetric channel, no single-particle contribution is  obtained so that only a continuum contribution results.



%
%
\end{document}